\documentclass[conference]{IEEEtran}
\usepackage{graphics}      
\usepackage[T1]{fontenc}   
\usepackage{txfonts}
\usepackage{mathptmx}
\usepackage[pdflang={en-US},pdftex]{hyperref}
\usepackage{color}
\usepackage{booktabs}
\usepackage{textcomp}
\usepackage{float}
\usepackage{enumitem}
\usepackage{hyperref}
\usepackage{cite}
\usepackage{amssymb,amsfonts}
\usepackage{algorithmic}
\usepackage{graphicx}
\usepackage{textcomp}
\usepackage{xcolor}
\usepackage{booktabs}
\usepackage{tabularx}
\usepackage{algorithm}
\usepackage{algorithmic}

\def\BibTeX{{\rm B\kern-.05em{\sc i\kern-.025em b}\kern-.08em
    T\kern-.1667em\lower.7ex\hbox{E}\kern-.125emX}}

\DeclareRobustCommand*{\IEEEauthorrefmark}[1]{%
\raisebox{0pt}[0pt][0pt]{\textsuperscript{\footnotesize #1}}%
}

\begin{document}

\title{Perfecting the Crime Machine \\
}

\author{\IEEEauthorblockN{Yigit Alparslan\IEEEauthorrefmark{1},
Ioanna Panagiotou\IEEEauthorrefmark{2}\, Willow Livengood\IEEEauthorrefmark{3},
Robert Kane\IEEEauthorrefmark{4}, Andrew Cohen\IEEEauthorrefmark{5}}
\IEEEauthorblockA{\IEEEauthorrefmark{1,}\IEEEauthorrefmark{2,}\IEEEauthorrefmark{3,}\IEEEauthorrefmark{5}Department of Electrical and Computer Engineering, Drexel University \\\IEEEauthorrefmark{4}Department of Criminology and Justice Studies, Drexel University\\
Philadelphia, PA\\
Email: \{ya332\IEEEauthorrefmark{1}, ip68\IEEEauthorrefmark{2}, wgl28\IEEEauthorrefmark{3}, rjk28\IEEEauthorrefmark{4}, arc334\IEEEauthorrefmark{5} \}@drexel.edu}}

\maketitle

\begin{abstract}
This study explores using different machine learning techniques and workflows to predict crime related statistics, specifically crime type in Philadelphia. We use crime location and time as main features, extract different features from the two features that our raw data has, and build models that would work with large number of class labels. We use different techniques to extract various features including combining unsupervised learning techniques and try to predict the crime type. Some of the models that we use are Support Vector Machines, Decision Trees, Random Forest, K-Nearest Neighbors. We report that the Random Forest as the best performing model to predict crime type with an error log loss of 2.3120.  
\end{abstract}


\section{Introduction}
Crime is a problem that we face every day in our society. Even though there are various reasons behind it, most of the reasons of crimes can be attributed to social-economical reasons. It is also shown that urban areas and cities show higher density of crime[1]. Crime also depends on different factors such as education, culture, economy level of neighbours and unemployment. There is a huge push towards using machine learning models to get statistics regarding crime predictions, to attest why they occur, when they would occur, and to whom it would occur[2][3][4][5][6][7]. One of the reasons we wanted to work with crime was because of individual incidents that we have seen on Drexel University campus, a rape incident dating around September 2019 that caused widespread backlash around Drexel University community and Philadelphia community regarding why Public Safety didn't take enough precautions. Philadelphia, being at the top 6 cities in the United States for population, and being our home appeals to us as a city that we can study, because we wanted to see if we could find any underlying reasons regarding crime by building predictive models, and see if we can systematically find those reasons with robust workflows. Some of the workflows that we adhere by in this study are feature extraction, model selection, parameter tuning for those models, and feature selection.

Studying crime patterns to prove hot spots would then bring attention to those areas. Once the crime statistics prove to be well-established, precautions such as increasing surveillance via facial recognition cameras can become a deterrant to criminals. Facial surveillance via street cameras on specifically chosen areas can be target by criminals. For this reason, it is important that, a database of records including video and audio can be saved to a remote server for further machine learning examinations to do facial recognition and classification as well as audio matching. Current literature has many examples of proving facial or audio recognition [14][15][16][17] to be robust with many techniques including adversarial attacks.

\section{Related Work}
There is a huge push towards building predictive models and fight against crime. Studies show that one of the techniques used widely in this crime field is to look at how dense the crime points are on a map. It has been shown that the existence of crime dense areas can be used as an indicator of the future crime areas since crime changes depend on several different reasons on a multidimensional layer, this has been widely accepted as an indicator of future crime. In this study, we wanted to differentiate ourselves by following approaches.

\begin{enumerate}
    \item Work with very large number of classes (30 labels)
    \item Create features that doesn't depend on the city
    \item Find optimal number of clusters in a data set 
    \item Cluster centers and use the distance as feature in our predictive models.
    \item Work with different supervised learning models that incorporate the aforementioned aspects hoping that it would increase our model accuracies.
\end{enumerate}

Researchers have focused on studying crime both from a time and location perspective[8]. The time perspective is the predictive aspect of crime as one might imagine. More specifically, one can create a grid on a city, and count the crime points on a grid and pose this problem as a regression over time series[9]. Other perspective is to use the location. Location might sound similar to the first time perspective but this is different and the difference lie on the fact that crime locations barely change over short amounts of time. So, if one were to study the crime dense neighbors of Philadelphia over a decade, and then guess the crime dense neighbors for the next year, month etc, one potential solution would be to flag the already existing crime dense areas and predict those neighbors as the future potential  crime dense areas. We have to realize that the literature uses a special word for this, that is crime hot spot. There are mathematical models that labels an area as crime hot spot or not based on a Euclidean distance, that is a linear kernel functions.

Even though current literature is built on top of these approaches, we want to remove the assumption that current literature has, even this meant deviating from the current literature approaches. For this reason, there is a narrow common ground between our findings and the common ground where we can compare our findings. This meant creating models that would not depend on the city. For example, as we have seen with the time perspective, a predictive model that poses this crime problem as a regression is a model that would need crime counts over time. To get the crime counts, most researches had to create grids on a city and count the crime counts for each grid and sum them over different periods of time[10][11]. This way, one can do regression single grid, and let's say, predict the crime count that one would expect in that crime grid cell for future time. One difficulty with that approach is that the data scientists or researches have to find a way to divide the city into different grid cells. This could be a problem since not every major city is a square, or has a compatible shape to be treated a rectangle/square. To mitigate this, we propose to create clusters in our data and use the clusters as a way of counting the crime instead of using a grid. More specifically, we create clusters by using our crime points for each year. We, then stack the clusters on top of each to get rid of the time element. Because we remove one of the constraints, now we can still get information from the time dimension that our data has without explicitly using it in our models. This approach of creating clusters removes the dependency of putting a grid into the city, and therefore it removes most of the preprocessing that a data scientist has to do to work with the crime data.

\section{Problem Statement}
Problems that we are tackling are as follows: 
\begin{enumerate}[label=(\roman*)]
    \item Can we predict crime type given location and time?
    \item Can we predict accurately if class number is very large? 
    \item Does incorporating features from unsupervised learning techniques improve our supervised models to predict crime type?
    \item Can we develop a systematic workflow to combine both learning (supervised/unsupervised) techniques for the crime data set that we work with?
\end{enumerate}

To answer these questions, we investigate creating clusters and looking at different supervised machine learning models.

\section{Basic Approach}
Our approach includes the following phases: 
\begin{enumerate}
    \item Data preprocessing, feature extraction
    \item Finding optimal number of clusters in our data set
    \item Creating Clusters for each year and stacking the cluster centers.
    \item Calculating the Euclidean distance from each crime point to cluster centers 
    \item Adding the distance features to previous, train different models including K-Nearest Neighbors, Logistic Regression, Decision Tree, Random Forest, Multi Layer Perceptron
\end{enumerate}

\begin{figure}[!ht]
\centering
  \includegraphics[width=0.9\columnwidth]{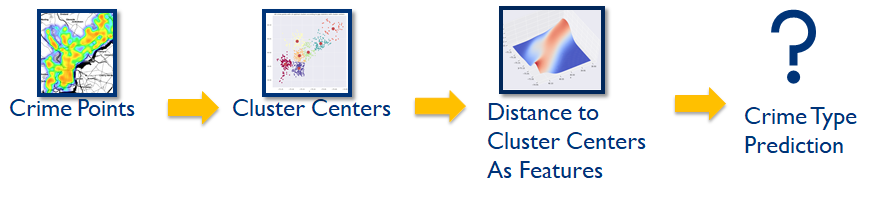}
  \caption{A snapshot of our workflow. First, we plot crime points, then we find optimal number clusters and then we find Euclidean distance from each crime point to its closest cluster center, which is then is used one of the features for our supervised learning models.}~\label{fig:workflow}
\end{figure}

We have two features in our data set. These features are mainly time and location. By using location and time, we can generate the following features via some data processing.
\begin{enumerate}
    \item Hour 
    \item Month
    \item Year
    \item DayOfWeek
    \item Is\_Weekend
    \item X 
    \item Y 
    \item Is\_Intersection
    \item Is\_Block
    \item Police District
    \item Street\_Type, (St, Blv, Ave etc)
\end{enumerate}

\begin{table}[htbp]
\centering
\caption{Raw features used in our data. We extracted more features from the ones that were presented to us in the raw data. Location helped us use Cartesian and polar coordinates together. Time helped us use day, hour, minute, is weekend, and day of week. }
\begin{tabularx}{\linewidth}{|l|l|l|X|}
\hline
\multicolumn{1}{|c|}{\textit{\textbf{X}}} & \multicolumn{1}{|c|}{\textit{\textbf{Y}}} & \multicolumn{1}{|c|}{\textit{\textbf{Date}}} &
\multicolumn{1}{c|}{\textit{\textbf{Description}}} \tabularnewline \hline
-75.174324 & 39.986978 & 4/3/2009 8:46 & Other Assaults \tabularnewline \hline
-75.238710 & 39.953566 & 2/2/2008 7:56 & Robbery Firearm \tabularnewline \hline
-75.069437 & 40.034939& 4/8/2007 2:54 & Driving Under Influence \tabularnewline \hline
-75.113286 & 39.996494 & 5/19/2006 11:37 & Thefts \tabularnewline \hline
-75.065362 & 40.046056 & 7/26/2006 13:35 & Other Assaults \tabularnewline \hline
\end{tabularx}
\label{table:dataset}   
\end{table}

\section{Data}
The data set used during this study has about 1.3 million samples. It has been collected by Open Data Philly City Council Organization[12]. For our supervised learning models, we used the 80/20 training set, we got about 838860 samples for training data and 262030 for testing data. This equates to using the first 9 years beginning from 2015 as training data and the remaining a year as the testing data set. The data set years ranged from 2006 to 2015. Some rows were missing some missing values. Missing values required us to do data pre-processing. In order to perform data processing, it is essential
to improve the data quality. There are a few techniques in practice, which are employed for the purpose
of data pre-processing. The techniques are data cleaning, feature selection, outlier detection, and component reduction and transformation. Before applying a classification algorithm usually some pre-processing is performed on the data set. Features are location and time for a crime points. Time for a crime point is dispatch time that the operator at 911 call center recorded. Therefore, the time is expressed with preciseness up until minute. The location for the crime point is the X, and Y coordinates of the crime point. Latitude measures angular distance from the equator to a point north or south of the equator. Longitude is an angular measure of east/west from the Prime Meridian, which has an angular measure of 0 since it is the beginning for that measure. Latitude values increase or decrease along the vertical axis, the Y axis. Longitude changes value along the horizontal access, the X axis. Philadelphia's latitude's range is slightly greater than its longitude range, which might make the Y feature more important. This can be easily seen with feature selection analysis that we did. More importantly, other city councils that gather the same type of crime data can easily see that the crime analysis is very tightly connected to the city structure and neighbors distribution. For this reason, data scientists and researches usually have to have prior knowledge regarding the physicality of the city. In our study, we tried to remove the assumptions regarding the city land space and focused on the features that can be generalized well such as dispatch time of crime, and angular measure of the crime point on Earth such as longitude and latitude.

We also look at individual classes to see the underlying patterns. Some patterns that we is that there are less crime during cold seasons than hot seasons over a year. Interestingly enough, in 2009, there is a dip in the number of classes that occurred per year.  Even though one might expect the otherwise situation since during recession when people panicked, one would have expected that there would be more crimes since people are more desperate. Around 6am is the safest hour in a day since most criminals are sleeping. Additionally, there is a peak in the crime count around lunch break. Since the data set is in Philadelphia, which is a highly populated urban city, there are more crimes in the lunch time compared to morning and slightly after lunch time.  Overall, crime count peaks in the evening between 8pm and 10pm and stays vrey high until 1am.
\begin{figure}[!ht]
\centering
  \includegraphics[width=0.9\columnwidth]{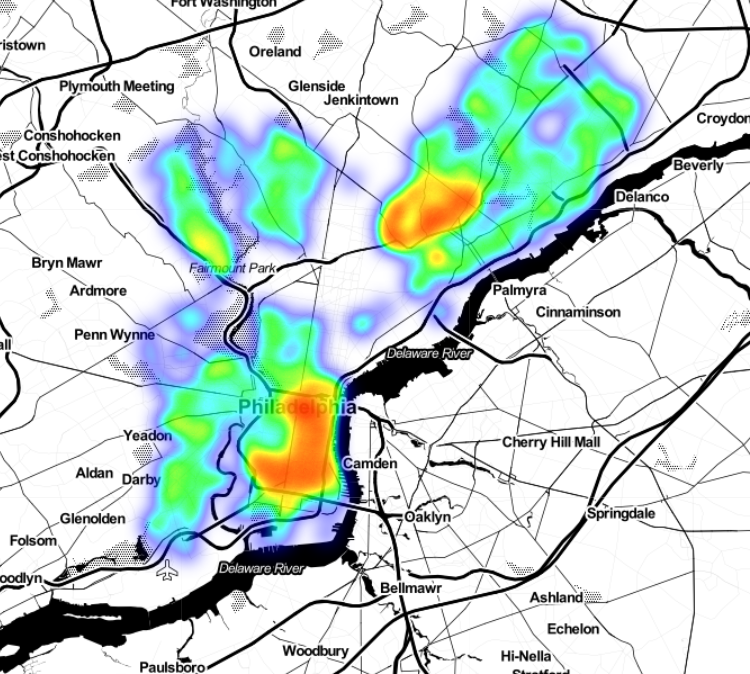}
  \caption{Crime points plotted for all years on Philadelphia map. Python's sci-kit library was used to plot the points}~\label{fig:crime_plot}
\end{figure}

\begin{figure}[!ht]
\centering
  \includegraphics[width=0.9\columnwidth]{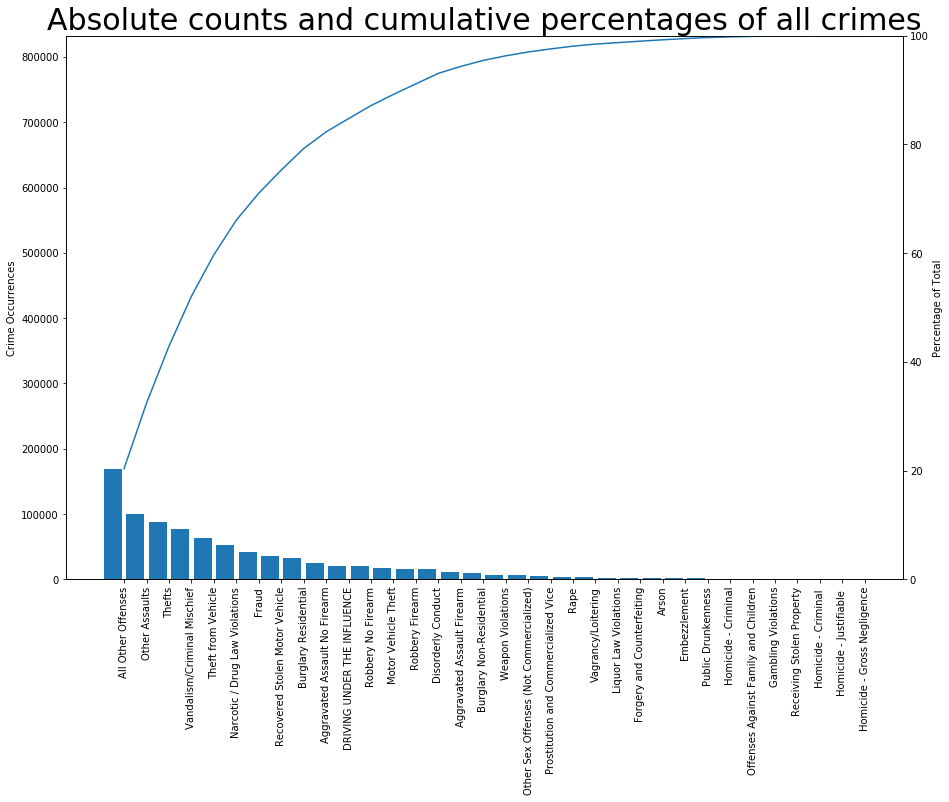}
  \caption{Crime counts for entire data set with crime labels. All the class labels can be seen on the horizontal axis. The vertical axis shows the running sum of those counts as percentage as well as absolute counts.}~\label{fig:absolute_counts}
\end{figure}

\begin{figure}[!ht]
\centering
  \includegraphics[width=0.9\columnwidth]{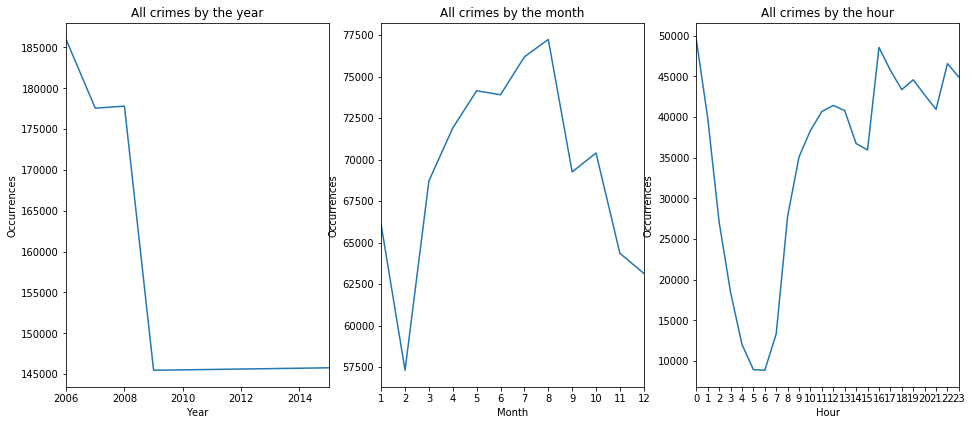}
  \caption{All crimes in the data set over years(i), months(i), and hour (iii) are shown. By plotting the all crime counts that occur over years, months, and hours we can see some of the underlying trends. We see that colder months, and early hours have less crime count. }~\label{fig:al_crimes_by_year}
\end{figure}

\begin{figure}[!ht]
\centering
  \includegraphics[width=0.9\columnwidth]{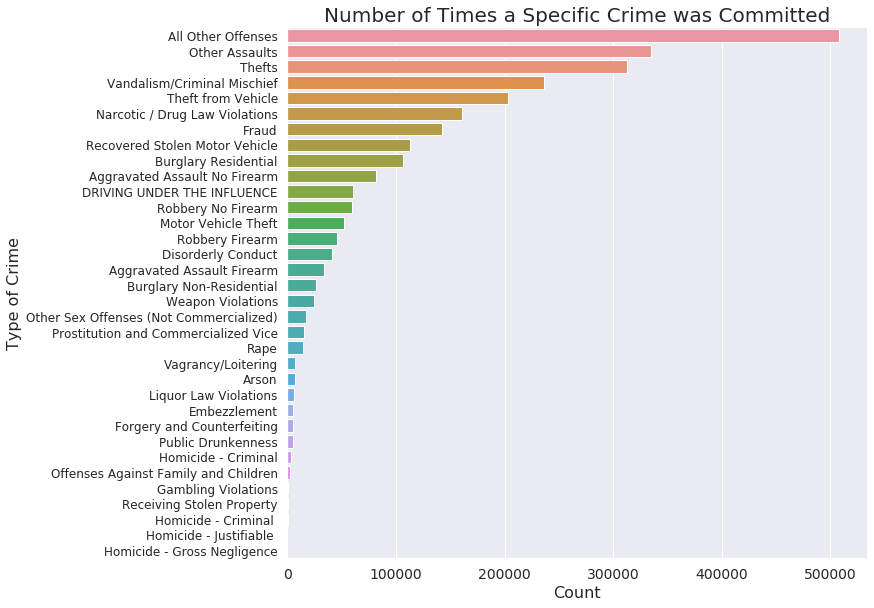}
  \caption{Distribution of crime counts over all class labels. There is a huge unbalanced crime count in the 'other' class label, which suggests that there might be some skew in our models.}~\label{fig:crime_dist}
\end{figure}

\begin{figure}[!ht]
\centering
  \includegraphics[width=0.9\columnwidth]{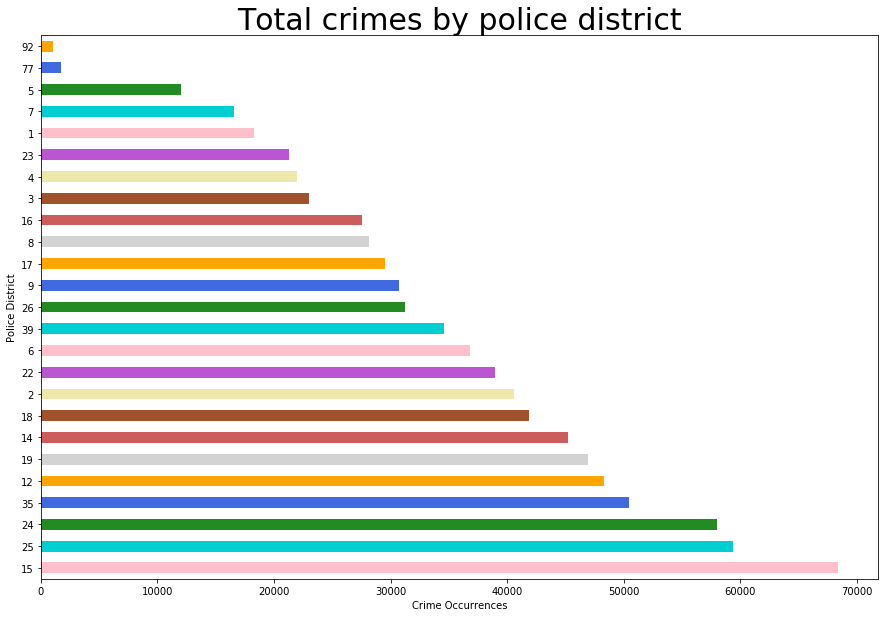}
  \caption{Total crimes by police district over Philadelphia police districts. There are about 30 police districts in Philadelphia and we see that the police districts in Center city, Northeast Philadelphia, and West Philadelphia frequent more crime than other police districts.}~\label{fig:total_crimes_wrt_police_district}
\end{figure}

Now, we look at some specific crime incidents and aggregate them over hours, months, and years. We see that some crime types such as prostitution and sex offenses occur very frequently during night time, and other crime types such as thefts and vandalism occur equally all day and remain stable in a day. We see that driving under influence occur at a very high rate between 10pm and 2 am, which is a natural time window for drivers who leave their parties after getting enough alcohol.

When we change the time scope from hours to months, we see that there is less crime incidents witnessed during cold months and that the hot months such as spring and summer see an increase in certain number of crime types such as thefts and prostitution.

If we look at the aggregation of crime types over years, we see that the trends get significantly harder to see. There are some general trends that we can mention. First, some crime types occur less over recent years such as vandalism. There are also some crime types that increase such as thefts. With thefts, we don't see a decline in the number of theft incidents.

\begin{figure}[!ht]
\centering
  \includegraphics[width=0.9\columnwidth]{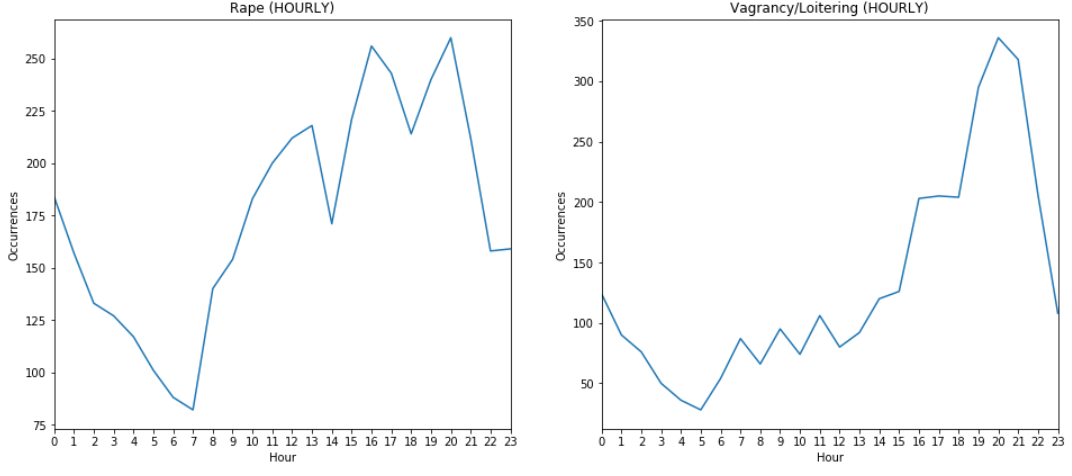}
  \caption{Rape(i) and Loitering(ii) crime counts over a day. We see that the lunchtime and evening hours, there is a peak.}~\label{fig:rape_hourly}
\end{figure}

\begin{figure}[!ht]
\centering
  \includegraphics[width=0.9\columnwidth]{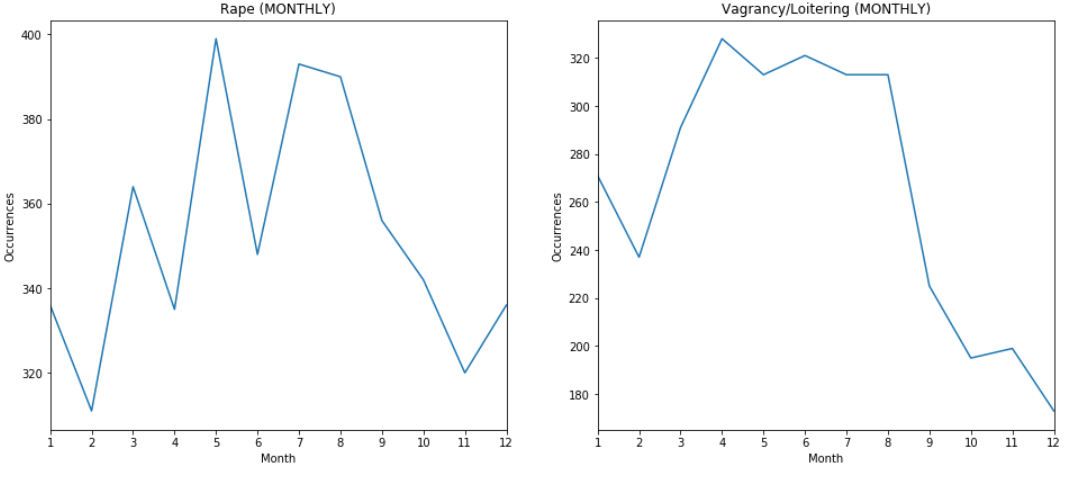}
  \caption{Rape(i) and Loitering(ii) crime counts over months. We see that the colder months witness less crime}~\label{fig:rape_monthly}
\end{figure}

\begin{figure}[!ht]
\centering
  \includegraphics[width=0.9\columnwidth]{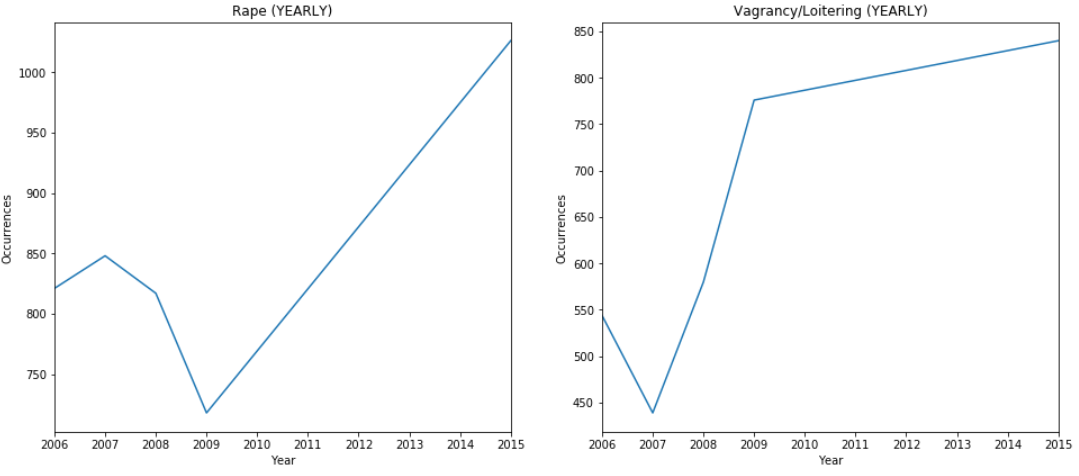}
  \caption{Rape(i) and Loitering(ii) crime counts over years. Looking at a year scale, it gets hard to see the underlying pattern, but we can say that the loitering decreases significantly during recent years.}~\label{fig:rape_yearly}
\end{figure}

\begin{figure}[!ht]
\centering
  \includegraphics[width=0.9\columnwidth]{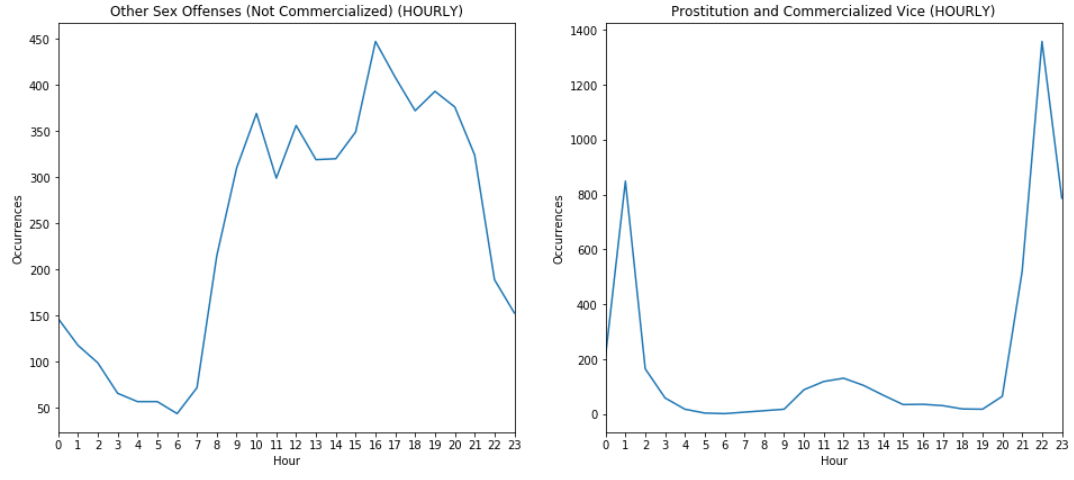}
  \caption{Other sex offenses not commercialized(i) and prostitution(ii) crime counts aggregated for a day. We see that there is an increase in the lunch time window for prostitution. }~\label{fig:other_sex_offenses_not_commercialized_hourly}
\end{figure}

\begin{figure}[!ht]
\centering
  \includegraphics[width=0.9\columnwidth]{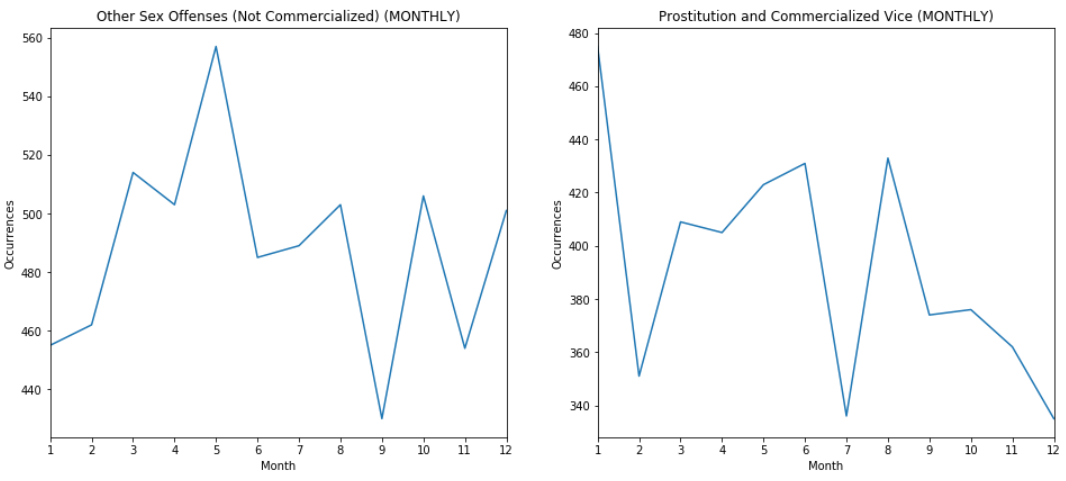}
  \caption{Other sex offenses not commercialized(i) and prostitution(ii) crime counts aggregated over months. We see that there is less crime in the fall and winter}~\label{fig:other_sex_offenses_not_commercialized_monthly}
\end{figure}

\begin{figure}[!ht]
\centering
  \includegraphics[width=0.9\columnwidth]{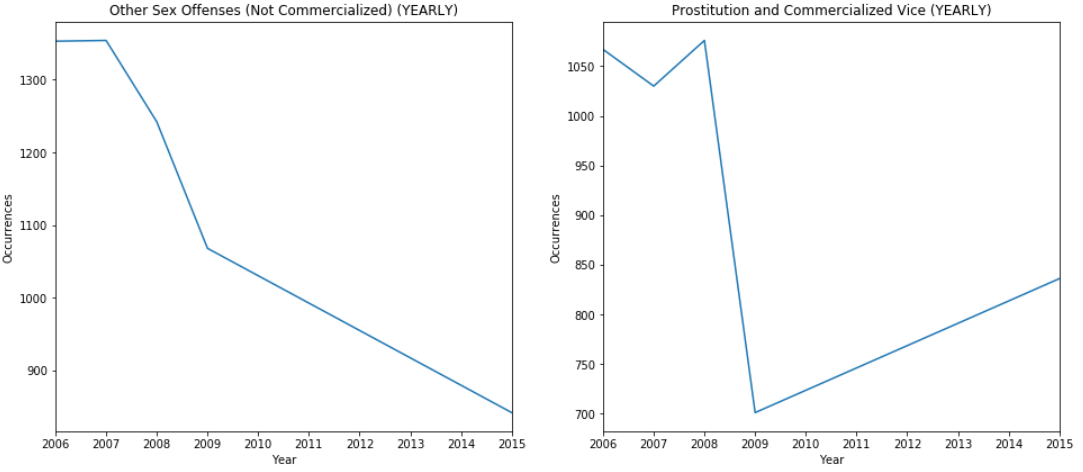}
  \caption{Other sex offenses not commercialized(i) and prostitution(ii) crime counts aggregated over a decade. It gets hard to see the underlying patterns when the scope is really zoomed out but when there is a general decline in the crime count for prostitution.}~\label{fig:other_sex_offenses_not_commercialized_yearly}
\end{figure}

\begin{figure}[!ht]
\centering
  \includegraphics[width=0.9\columnwidth]{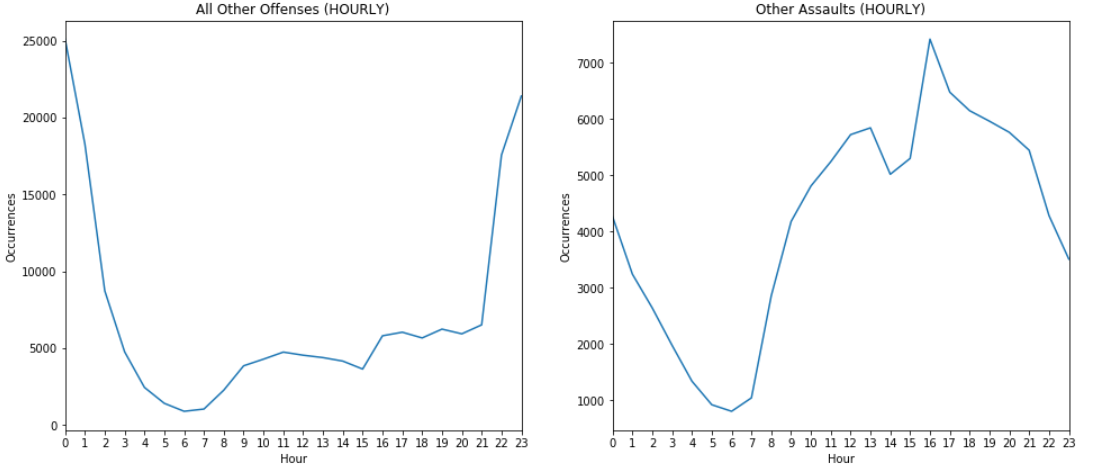}
  \caption{All other offenses(i) and other assaults(ii) crime counts aggregated for a day. We see that there are more crime incidents in the evening hours }~\label{fig:all_other_offenses_hourly}
\end{figure}

\begin{figure}[!ht]
\centering
  \includegraphics[width=0.9\columnwidth]{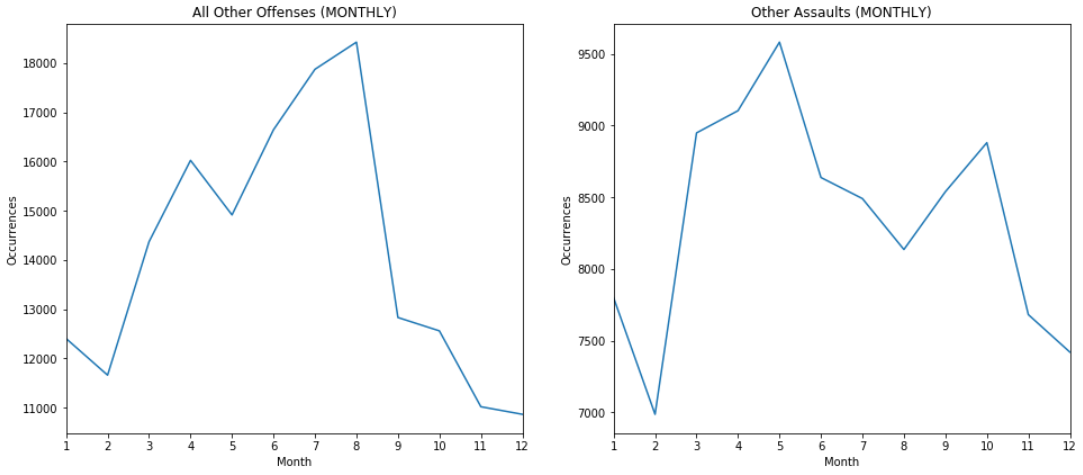}
  \caption{All other offenses(i) and other assaults(ii) crime counts aggregated over months. We see that there are less crime crime incidents over cold months such as winter and fall seasons.}~\label{fig:all_other_offenses_monthly}
\end{figure}

\begin{figure}[!ht]
\centering
  \includegraphics[width=0.9\columnwidth]{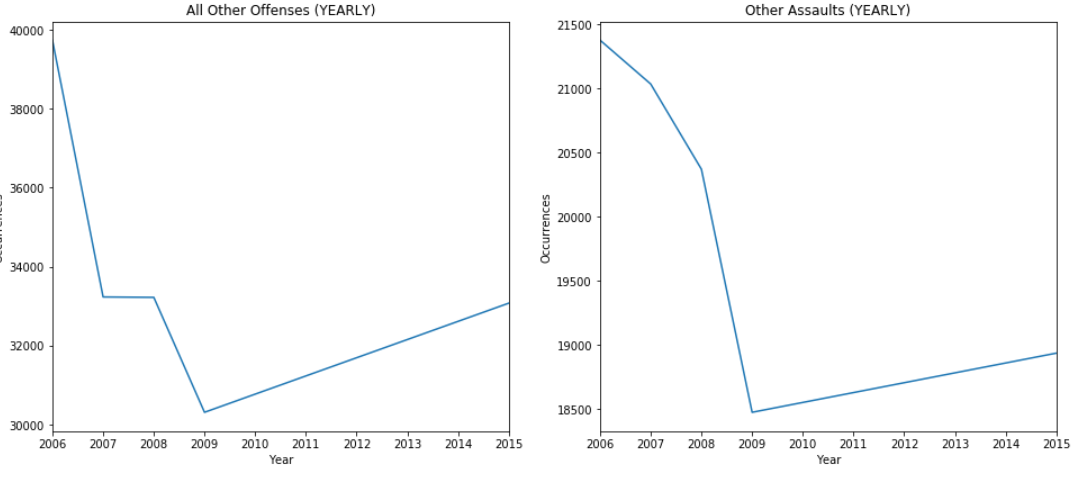}
  \caption{All other offenses(i) and other assaults(ii) crime counts aggregated over years. We see that there is a decline in the crime counts over recent years.}~\label{fig:all_other_offenses_yearly}
\end{figure}

\begin{figure}[!ht]
\centering
  \includegraphics[width=0.9\columnwidth]{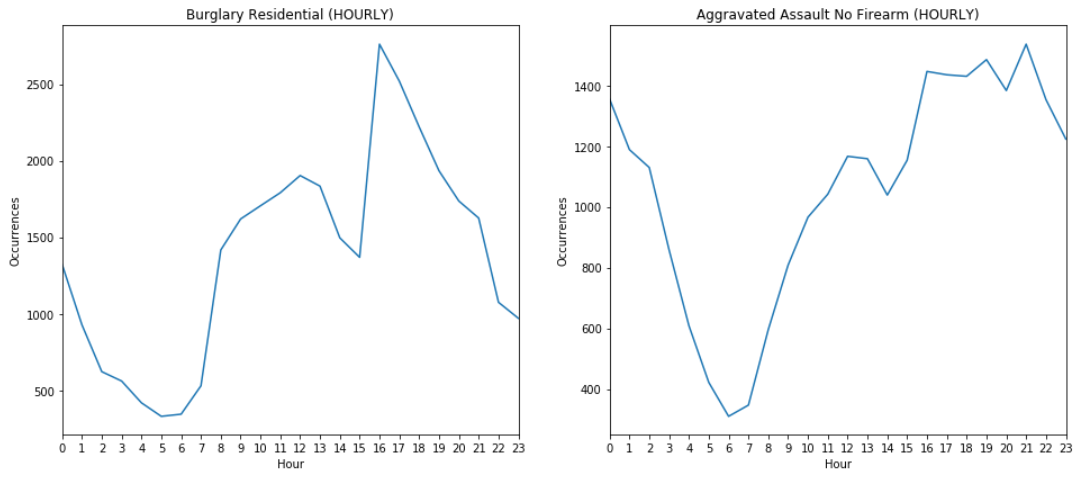}
  \caption{Residential burglary(i) and aggravated assaults with no firearms(ii) crime counts aggregated over hours. We see that there is a huge jump in the evening hours.}~\label{fig:burglary_residential_hourly}
\end{figure}

\begin{figure}[!ht]
\centering
  \includegraphics[width=0.9\columnwidth]{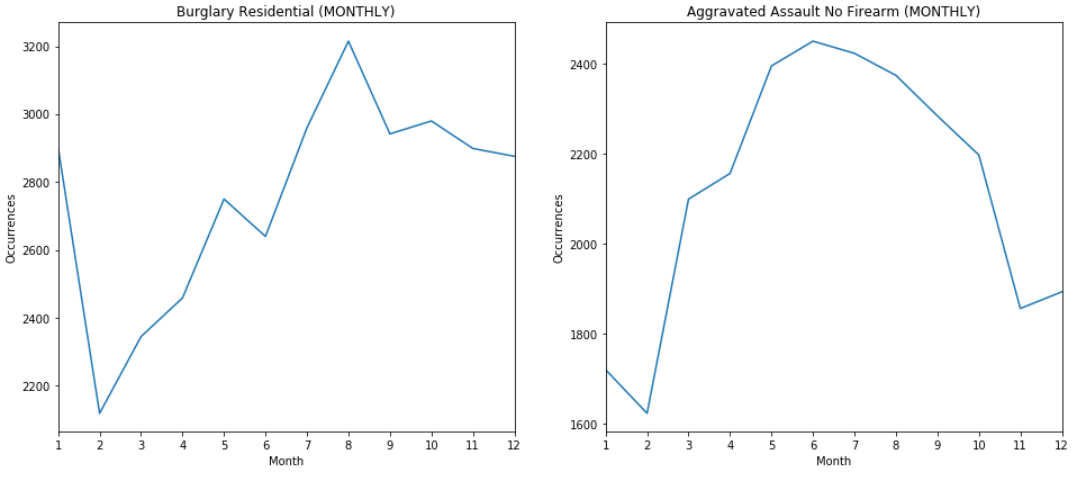}
  \caption{Residential burglary(i) and aggravated assaults with no firearms(ii) crime counts aggregated over months. We see that there are less crime crime incidents over cold months such as winter and fall seasons.}~\label{fig:burglary_residential_monthly}
\end{figure}

\begin{figure}[!ht]
\centering
  \includegraphics[width=0.9\columnwidth]{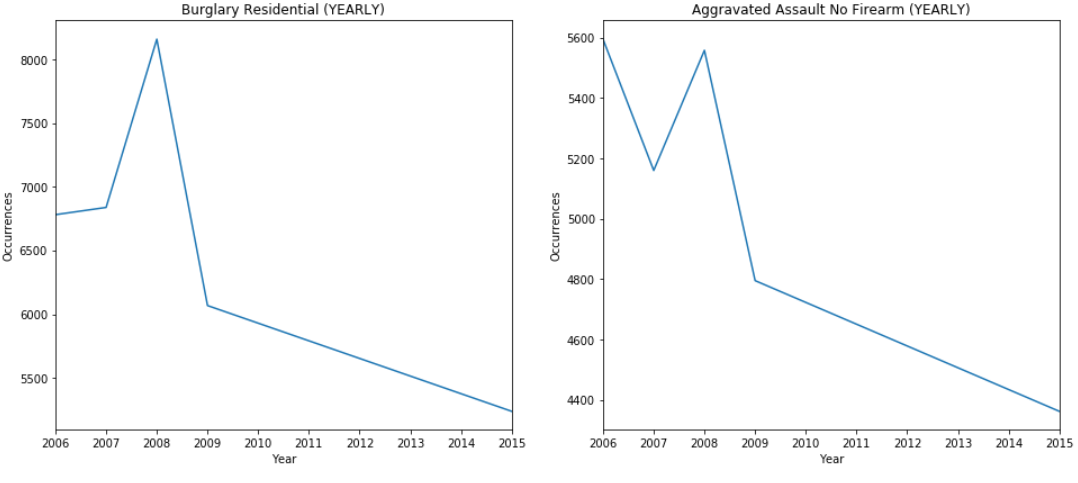}
  \caption{Residential burglary(i) and aggravated assaults with no firearms(ii) crime counts aggregated over years. We see that there are less crime crime incidents over recent years, but when the scope is years, it gets hard to tell the underlying patterns}~\label{fig:burglary_residential_yearly}
\end{figure}

\begin{figure}[!ht]
\centering
  \includegraphics[width=0.9\columnwidth]{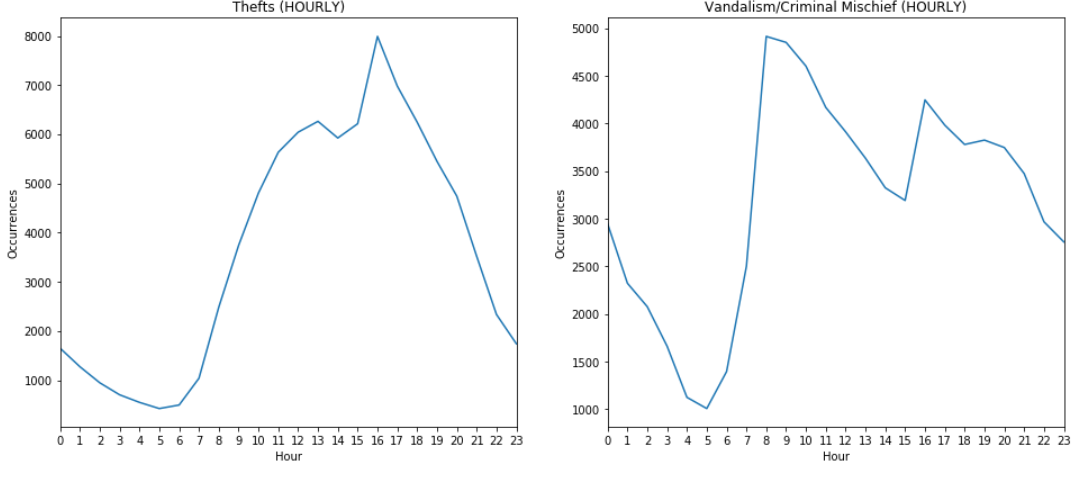}
  \caption{Thefts(i) and Vandalism(ii) crime counts aggregated over hours. We see that theft occurs all the time when humans are awake, and vandalism peaks significantly during daylight.}~\label{fig:thefts_hourly}
\end{figure}

\begin{figure}[!ht]
\centering
  \includegraphics[width=0.9\columnwidth]{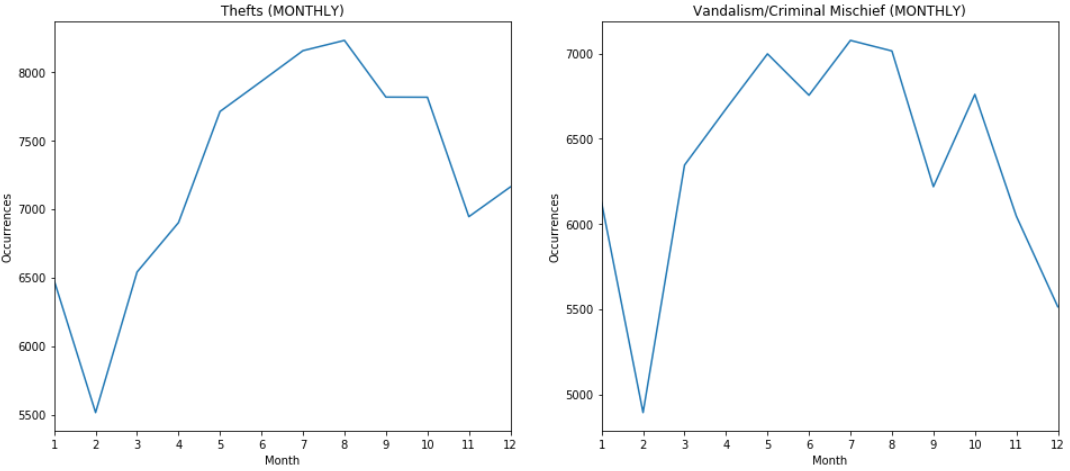}
  \caption{Thefts(i) and Vandalism(ii) crime counts aggregated over months. We see that cold months witness less crime incidents.}~\label{fig:thefts_monthly}
\end{figure}

\begin{figure}[!ht]
\centering
  \includegraphics[width=0.9\columnwidth]{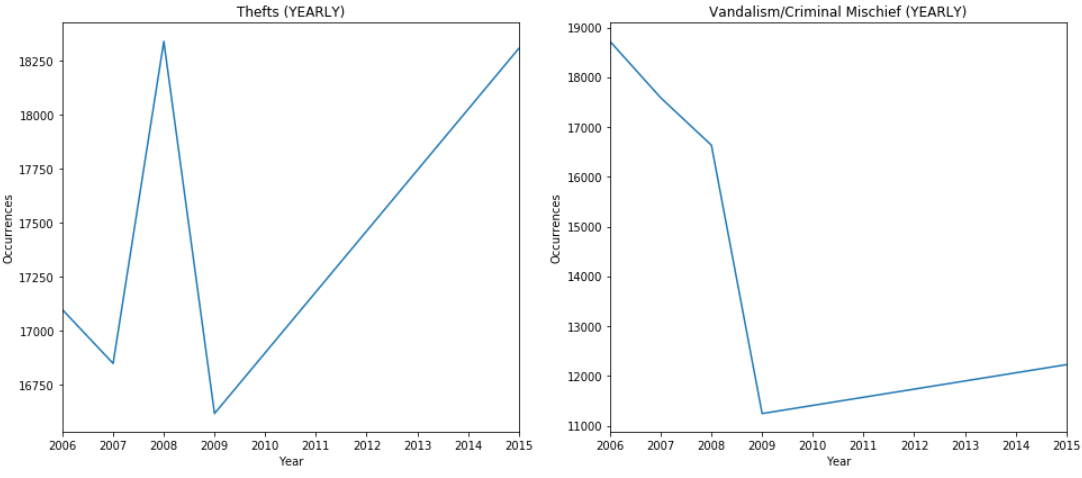}
  \caption{Thefts(i) and Vandalism(ii) crime counts aggregated over years. Year score is making the underlying patterns hard to see, but we witness less crime vandalism occurring n the recent years. Thefts seem to be be increasing.}~\label{fig:thefts_yearly}
\end{figure}

\section{Experiment Results, Analysis and Performance Evaluation}

Quick Summary for results are as follows:
\begin{enumerate}
    \item Random Forest is most sensitive to the minute and the hour. 
    \item Random Forest is the best performing model, which aligns with the current literature. 
    \item Support Vector Machines over 30 labels fails to run to completion in Google Cloud Compute Engine Service.
    \item The optimal number of clusters for all years is7 but when we take each year as a separate data set, we see that the optimal number of clusters varies between 7 and 10.
    \item Bayesian Inference works significantly well with 30 class labels, achieving around 27\% mean accuracy compared to logistic regression with 5\%and K Nearest Neighbors with 19\% accuracies.
\end{enumerate}

Having said all these results, now we expand upon them with details here.
\subsection{Unsupervised Learning}
Unsupervised learning techniques are methods where one employs systematic methods to a data set without any labels to understand the underlying features regarding the data. The techniques that we use in this study is K-Means clustering algorithm. One really important aspect about this method is to find the optimal number of clusters in one's data set. For this objective, we employed two different methods, namely elbow method and gap statistics. 
Elbow method can be employed like this:
\begin{algorithm}
\caption{Calculate optimal number of clusters - Elbow Method}
\begin{algorithmic}
\STATE For each k value, we initialised k-means and used the inertia attribute to identify the sum of squared distances of samples to the nearest cluster centre.
\STATE As k increases, the sum of squared distance tends to zero. Imagine we set k to its maximum value n (where n is number of samples) each sample will form its own cluster meaning sum of squared distances equals zero.
\STATE If the plot looks like an arm, then the elbow on the arm is optimal k.

\end{algorithmic}
\end{algorithm}

\begin{algorithm}
\caption{Calculate optimal number of clusters - Gap Statistics}
\begin{algorithmic}
\STATE Cluster the observed data, varying the number of clusters from k = 1, …, kmax, and compute the corresponding total within intra-cluster variation Wk.
\STATE Generate B reference data sets with a random uniform distribution. 
\STATE Cluster each of these reference data sets with varying number of clusters k = 1, kmax, and compute the corresponding total within intracluster variation Wkb.
\STATE Compute the estimated gap statistic as the deviation of the observed Wk value from its expected value Wkb under the null hypothesis: Gap(k)=$\sum_{b=1}^{B} log(W\textsubscript{kb})$ $-$ log(W\textsubscript{k}).
\STATE Compute also the standard deviation of the statistics.
\STATE Choose the number of clusters as the smallest value of k such that the gap statistic is within one standard deviation of the gap at k+1: Gap(k)$\geq$Gap(k+1)$-$ --s\textsubscript{k+1} where s is the standard deviation.

\end{algorithmic}
\end{algorithm}

\begin{figure}[!ht]
\centering
  \includegraphics[width=0.9\columnwidth]{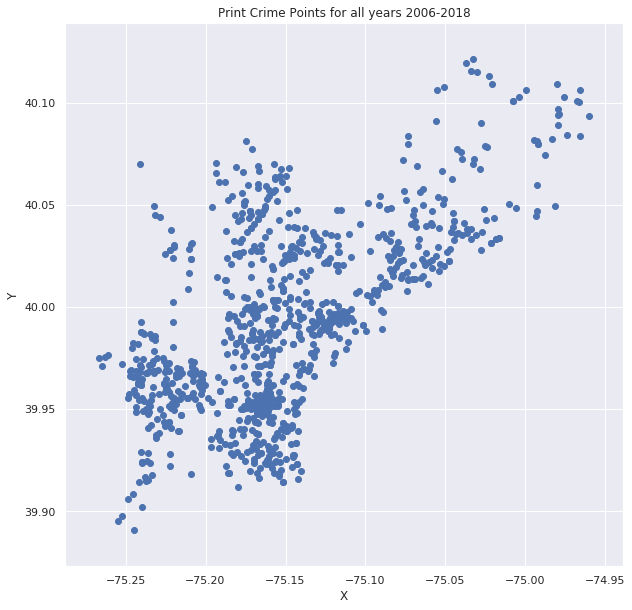}
  \caption{Plot of all crime for 2006-2018. X and Y coordinates on a Cartesian's coordinate system can be used to plot all the crime points in this map without the explicit Philadelphia borders.}~\label{fig:figure13}
\end{figure}

\begin{figure}[!ht]
\centering
  \includegraphics[width=0.9\columnwidth]{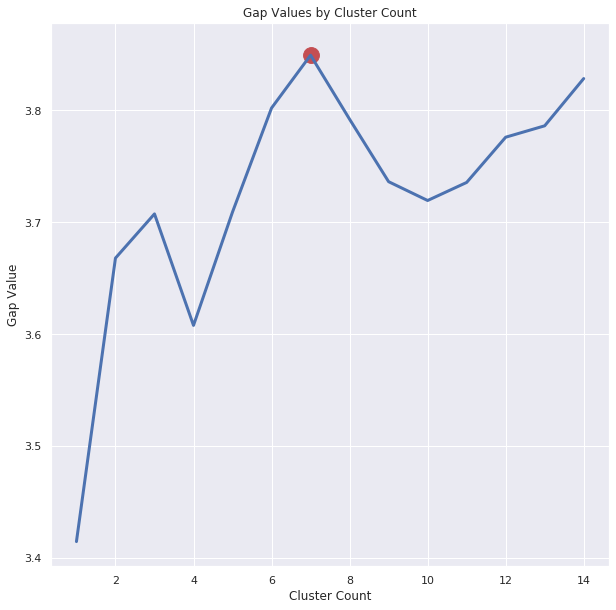}
  \caption{Gap Values by Crime count over a range to determine the optimal cluster count. We use the Gap statistics over the range of 1 to 16 and pick the index of the highest gap value as the optimal cluster count. This method gives 7 as the optimal cluster number}~\label{fig:gap_values_by_cluster_count}
\end{figure}

 \begin{figure}[!ht]
\centering
  \includegraphics[width=0.9\columnwidth]{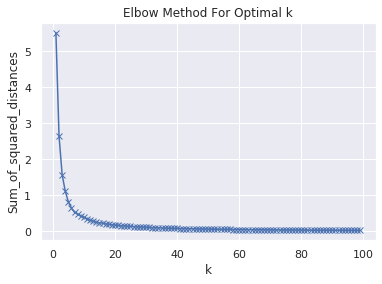}
  \caption{Elbow Method to determine the optimal cluster values. This method gives 3 as the optimal cluster count.}~\label{fig:elbow_method}
\end{figure}
Gap statistics method gives 7 as the optimal cluster number when applied to all the years in our data set. Elbow method gives 3 as the optimal cluster number when applied to all years in our data set. Because we have about 1 million data points, we want to maximize the variance on distances that we calculate to the cluster centers, and go with the k=7 optimal cluster count. We apply the K-Means clustering algorithm and see the results in Figure \ref{fig:cluster_centers}

\begin{figure}[!ht]
\centering
  \includegraphics[width=0.9\columnwidth]{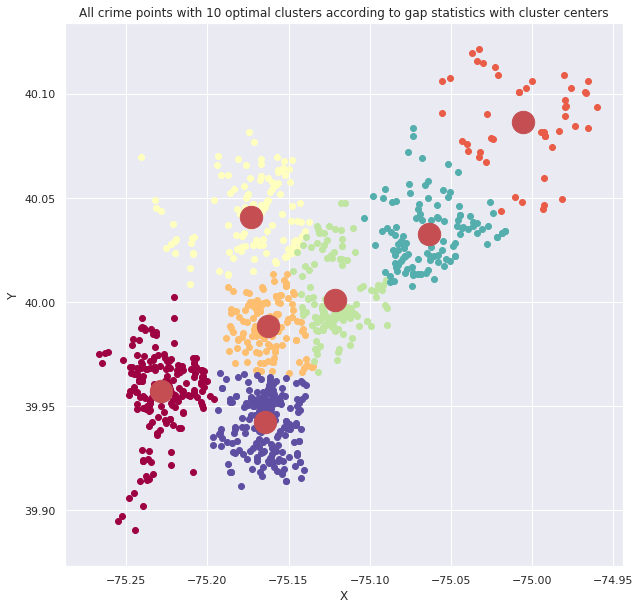}
  \caption{Average similarity distance when PCA is applied to all 25 adversarial examples with respect to percentage of components used to craft the adversarial effect during PCA}~\label{fig:gap}
\end{figure}

We plot only the cluster centers for each year and show the results in Figure 28. Cluster centers align on the direction from west to North East indicating that the crime points are gathered around the West Philadelphia and North East Philadelphia areas. To see the effect of cluster centers on the map, we apply a Gaussian Density Function to draw contours where height of the surface indicates the density of the crime happening in the future. This aspect really resembles the approach that we mentioned in our Related Work section when we introduced the concepts that literature took in this field.One being the time aspect, and one being the location aspect. Northeast and West Philadelphia achieved the tallest surface heights.

\begin{figure}[!ht]
\centering
  \includegraphics[width=0.9\columnwidth]{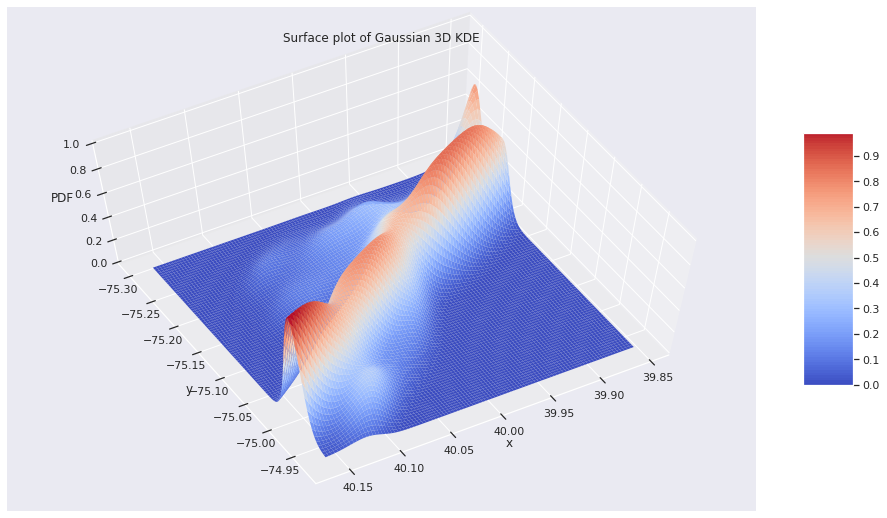}
  \caption{Average similarity distance when PCA is applied to all 25 adversarial examples with respect to percentage of components used to craft the adversarial effect during PCA}~\label{fig:3d_kernel}
\end{figure}

\begin{figure}[!ht]
\centering
  \includegraphics[width=0.9\columnwidth]{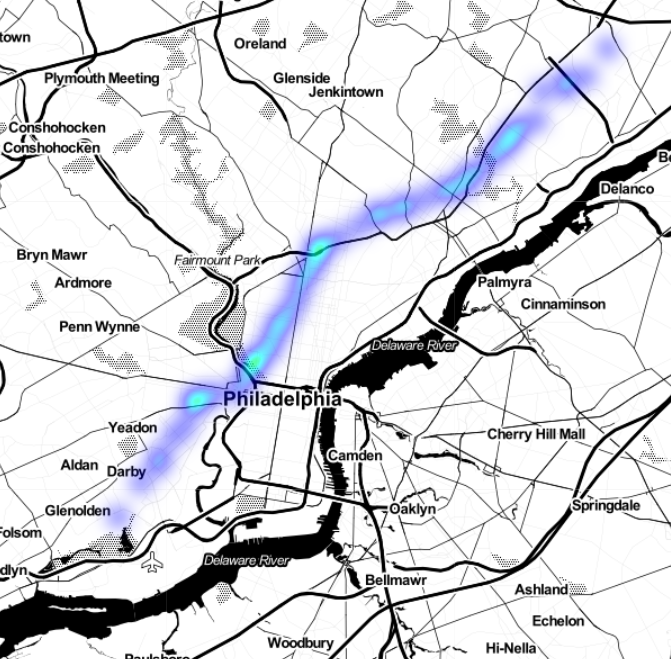}
  \caption{Average similarity distance when PCA is applied to all 25 adversarial examples with respect to percentage of components used to craft the adversarial effect during PCA}~\label{fig:cluster_centers}
\end{figure}

 \subsection{Supervised Learning}
Supervised learning methods are techniques that employ training data, a cost function and a testing data where training data is used to fit a data and testing data is used to report how well the the fit was behaved. Out of 1.3  millions samples in the training data set. We got about 838860 samples for training data  and  262030  for  testing  data.  This  equates  to  using  the first  9  years  beginning  from  2015  as  training  data  and  the remaining years 2015, 2016, 2017 and 2018  years  as  the  testing  data  set.  Some  rows  had  missing  missing  values.  Missing  values  required  us  to  do  data  pre-processing and drop them.The class labels can be seen in Table \ref{table:Labels}. The models that we employed in order to predict the crime types are as follows:
\begin{enumerate}
    \item K Nearest Neighbors
    \item Naive Bayesian Inference
    \item Decision Tree
    \item Random Forest
    \item Logistic Regression
    \item Support Vector Machine
    \item Multi Layer Perceptron
\end{enumerate}
We choose KNN with 5 neighbors. KNN is a classifier that makes the classification output based on the majority of votes of the k nearest neighbors.Naive Bayes methods naively employs inference by assuming that the feature pairs are independent.
We also used Decision Tree with a confidence factor 0.3. Decision Trees are supervised learning models that achieves the value of the target variable by learning simple splitting rules/decision rules on the data set.
Random Forest that we used had 10 trees. A random forest is a model that combines several decision trees on several sub-samples of the data set and use the averaging to improve the predictive accuracy. Since it uses several trees, it is also expected to generalize well and avoid over fitting.
Logistic Regression is a supervised learning method which is well suited to be a binary classifier and can also be used for multi class classification problems. It uses a log function in order to produce probability values over classes which then can be used to predict classes. 
Support Vector Machines (SVM) are supervised learning machines. They implement a good generalization on a limited number of learning
patterns inferred based on the features that we used. It uses a linear kernel and tries to separate the crime points in a very high dimensional space which is likely to have a linear hyper plane.
Multi Layer Perceptrons are layered supervised learning models that tries to find a hyper plane in order to separable the data. We employed one hidden layer of 150 neurons.

During our study, we employed a free Google Cloud Compute Engine Service with free 12 hours of GPU access in order to take advantage of fast cloud computing. We encourage the reader to see the specifications here[12]. We had several time outs whilst training the Support Vector Machines and the Multi Layer Perceptron models, and therefore, we don't report their results in this study. 
\begin{table}[htbp]
\centering
\begin{tabularx}{\linewidth}{|l|X|}
\hline
\multicolumn{1}{|c|}{\textit{\textbf{Class Label Index}}} & \multicolumn{1}{c|}{\textit{\textbf{Class Labels Used in the Supervised Models}}} \tabularnewline \hline
0 &Aggravated Assault Firearm\tabularnewline \hline
1 &Aggravated Assault No Firearm\tabularnewline \hline
2 &All Other Offenses \tabularnewline \hline
3 &Arson \tabularnewline \hline
4 &Burglary Non-Residential \tabularnewline \hline
5 &Burglary Residential \tabularnewline \hline
6 &Driving Under Influence \tabularnewline \hline
7 &Disorderly Conduct \tabularnewline \hline
8 &Embezzlement \tabularnewline \hline
9 &Forgery and Counterfeiting \tabularnewline \hline
10 &Fraud \tabularnewline \hline
11 &Gambling Violations \tabularnewline \hline
12 &Homicide - Criminal  \tabularnewline \hline
13 &Homicide - Gross Negligence \tabularnewline \hline
14 &Homicide - Justifiable  \tabularnewline \hline
15 &Liquor Law Violations \tabularnewline \hline
16 &Motor Vehicle Theft \tabularnewline \hline
17 &Narcotic / Drug Law Violations \tabularnewline \hline
18 &Offenses Against Family and Children \tabularnewline \hline
19 &Other Assaults \tabularnewline \hline
20 &Other Sex Offenses (Not Commercialized) \tabularnewline \hline
21 &Prostitution and Commercialized Vice \tabularnewline \hline
22 &Public Drunkenness \tabularnewline \hline
23 &Rape \tabularnewline \hline
24 &Receiving Stolen Property \tabularnewline \hline
25 &Recovered Stolen Motor Vehicle \tabularnewline \hline
26 &Robbery Firearm \tabularnewline \hline
27 &Robbery No Firearm \tabularnewline \hline
28 &Theft from Vehicle \tabularnewline \hline
29 &Thefts \tabularnewline \hline
30 &Vagrancy/Loitering \tabularnewline \hline
31 &Vandalism/Criminal Mischief \tabularnewline \hline
32 &Weapon Violations \tabularnewline \hline

\end{tabularx}
\caption{Class Labels that were used for the supervised learning models. Labels were given an integer to make the plot and training easier.}
\label{table:Labels}   
\end{table}

All the models were done by using Python's sci-kit library and the preprocessing was done by first reading from excel file, splitting it into two: first for training and second for testing with 80\% and 20\% ratios respectively.

\begin{figure}[!ht]
\centering
  \includegraphics[width=0.9\columnwidth]{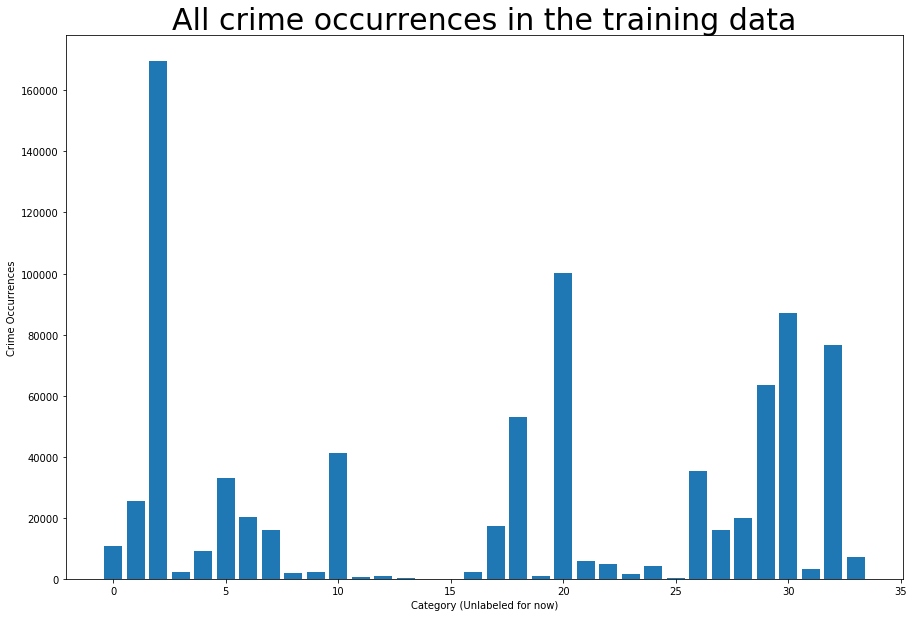}
  \caption{All crime occurrences in the training data. Crime occurrences in the training data are seen over the class labels, which are encoded as integers. We see that some classes have much greater crime count compared to others. }~\label{fig:crime_occurrences_in_training_data}
\end{figure}

\begin{figure}[!ht]
\centering
  \includegraphics[width=0.9\columnwidth]{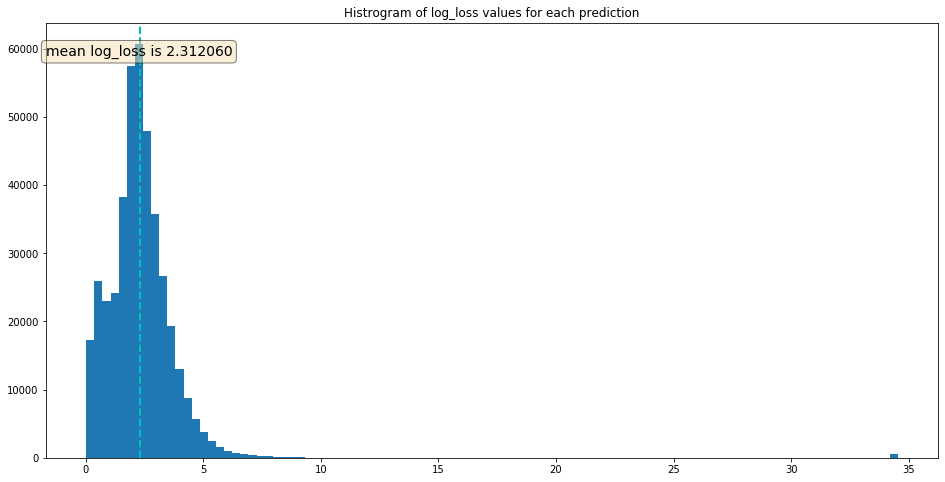}
  \caption{Histogram of different class labels and log loss values after predicting each label. Mean log loss gives us how well we are doing collectively and individually when we are predicting the crime types. }~\label{fig:mean_log}
\end{figure}
 
 We also take a look at the number components that we can keep the high variance. This can help us eliminate components that don't give extra information, or important information.
 
\begin{figure}[!ht]
\centering
  \includegraphics[width=0.9\columnwidth]{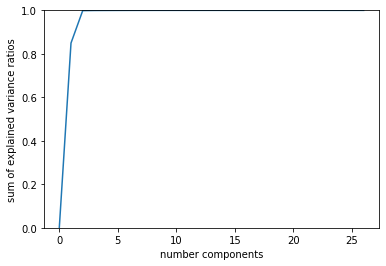}
  \caption{Principal component analysis over features. Ranking features and keeping the ones tat have the highest information gain can be achieved with applying principal component analysis.}~\label{fig:pca}
\end{figure}

When we apply Principal component analysis, we see that applying PCA to our model will decrease the performance, This can be attributed to the fact that we are working very small number features and  because essential information is lost in the PCA process, we lose information immediately after we start applying PCA.

For each row, a uniform probability prediction (no machine learning required), where each label has a 1/34 probability would give a log loss score of:  
$$
log loss = -\log(\frac{1}{34})= 3.5263605
$$
So if we calculate the log loss score per label, we can see that for what labels, we are performing worse than the base line probability.

\begin{figure}[!ht]
\centering
  \includegraphics[width=0.9\columnwidth]{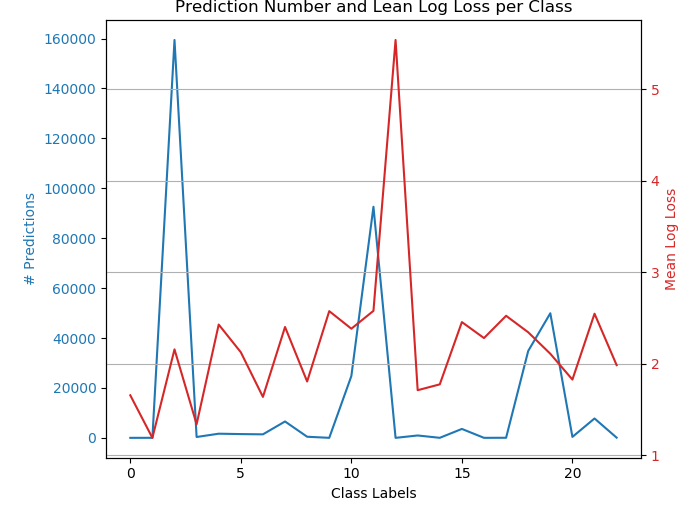}
  \caption{Distribution of predictions over all classes. We see that our Random Forest is not doing well for classes that have a very large number of sample counts compared to classes that have less counts.}~\label{fig:predvsmeanlogll}
\end{figure}

\begin{figure}[!ht]
\centering
  \includegraphics[width=0.9\columnwidth]{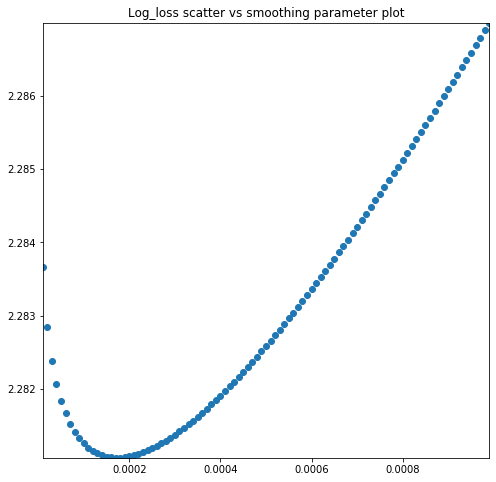}
  \caption{Smoothing probability values over all class labels. We get the lowest $log_loss$ score of: 2.281062, with smoothing parameter: 0.000170. The improvement is 0.169367\%}~\label{fig:smoothing}
\end{figure}

As it can be seen in Figure \ref{fig:smoothing}, we apply a smoothing parameter in order to improve the accuracy of our models.We add a small value to all the probability predictions. This is to achieve that we don't have any 0 value probability.   
Note that while the $\sum{row_{predictions}} > {1}$ for each row of the prediction matrix, this is not an issue. $Log_loss$ function used by Python rescales the matrix back to $\sum{row_{predictions}} = 1$. 

 Figure \ref{fig:smoothing}, gives insight to adding a smoothing parameter to the probability predictions over different classes. We get the lowest $log_loss$ score of: 2.281062, with smoothing parameter: 0.000170. The improvement is 0.169367\%

\begin{table}[htbp]
\caption{Distribution of mispredictions over all classes and the mean log loss of the mispredictions}
\centering
\begin{tabularx}{\linewidth}{|l|l|X|}
\hline
\multicolumn{1}{|c|}{\textit{\textbf{Label}}} & \multicolumn{1}{|c|}{\textit{\textbf{No of Mispredictions}}} & \multicolumn{1}{c|}{\textit{\textbf{Mean Log Loss}}} \tabularnewline \hline
0 & 4  & 1.654153 \tabularnewline \hline
1 & 18  & 1.918882\tabularnewline \hline
2 & 159416  & 2.156223\tabularnewline \hline
4 & 344  & 1.337823\tabularnewline \hline
5 & 1663  & 2.427808\tabularnewline \hline
6 & 1510  & 2.124929\tabularnewline \hline
7 & 1409  & 1.636606\tabularnewline \hline
10 & 6561  & 2.401358\tabularnewline \hline
16 & 446  & 1.805052\tabularnewline \hline
17 &   7  & 2.574474\tabularnewline \hline
18 & 24767  & 2.382460\tabularnewline \hline
20 & 92597  & 2.577907\tabularnewline \hline
21 &   1  & 5.540110\tabularnewline \hline
22 & 952  & 1.709928\tabularnewline \hline
23 &  15  & 1.773899\tabularnewline \hline
26 & 3573  & 2.454954\tabularnewline \hline
27 &   9  & 2.280118\tabularnewline \hline
28 &  34  & 2.523617\tabularnewline \hline
29 & 34812  & 2.343397\tabularnewline \hline
30 & 49930  & 2.108511\tabularnewline \hline
31 & 386  & 1.826013\tabularnewline \hline
32 & 37763  & 2.546528\tabularnewline \hline
33 &  79  & 1.984121\tabularnewline \hline
\end{tabularx}
\label{table:mispredictions_over_classes}   
\end{table}

\begin{table}[htbp]
\caption{Feature rankings and Weights. Looking at features contribution to the predictions can tell us which feature to focus on during our predictions. For this ranking, we see that the X, Y and other time location features such as minute and hour were the features that had the most value for our predictions.}
\centering
\begin{tabularx}{\linewidth}{|l|l|X|}
\hline
\multicolumn{1}{|c|}{\textit{\textbf{Rank}}} & \multicolumn{1}{|c|}{\textit{\textbf{Feature}}} & \multicolumn{1}{c|}{\textit{\textbf{Weight}}} \tabularnewline \hline
0 & Hour Zone & 0.091309\tabularnewline \hline
1 & Hour & 0.089656\tabularnewline \hline
2 & Y & 0.062149\tabularnewline \hline
3 & Rot60X & 0.060207\tabularnewline \hline
4 & Radius & 0.058731\tabularnewline \hline
5 & Rot45X & 0.057662\tabularnewline \hline
6 & Angle & 0.057590\tabularnewline \hline
7 & X & 22 0.05119\tabularnewline \hline
8 & Rot30Y & 0.056596\tabularnewline \hline
9 & Rot30X & 0.056554\tabularnewline \hline
10 & Rot60Y & 0.055187\tabularnewline \hline
11 & Rot45Y & 0.054602\tabularnewline \hline
12 & Street1 & 0.0038214\tabularnewline \hline
13 & Minute & 0.032564\tabularnewline \hline
14 & WeekOfYear & 0.031246\tabularnewline \hline
15 & Year & 0.028269\tabularnewline \hline
16 & Day & 0.02737\tabularnewline \hline
17 & DayOfWeekNum & 0.019447\tabularnewline \hline
18 & PdDistrictNum & 0.017935\tabularnewline \hline
19 & Month & 0.017296\tabularnewline \hline
20 & Street2 & 0.010806\tabularnewline \hline
21 & Season & 0.008322\tabularnewline \hline
22 & IsWeekend & 0.007317\tabularnewline \hline
23 & IsIntersection & 0.003908\tabularnewline \hline
24 & StreetType & 0.000005\tabularnewline \hline
25 & IsBlock & 0.00000\tabularnewline \hline
\end{tabularx}
\label{table:feature_ranking}   
\end{table}

Some of the feature rankings that we have done can be seen in Table \ref{table:feature_ranking}
\begin{figure}[!ht]
\centering
  \includegraphics[width=0.9\columnwidth]{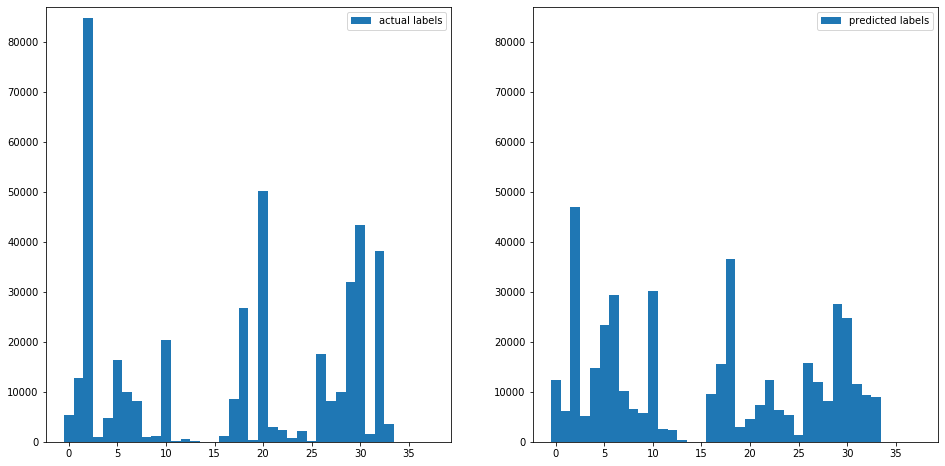}
  \caption{Actual vs Predicted Labels. Actual labels are the true integers that represent the crime type. Predicted labels for Random Forest are the classification outputs by the Random Forest.}~\label{fig:actual_vs_predicted_labels}
\end{figure}

\begin{table}[htbp]
\caption{Model Performance Metrics }
\centering
\begin{tabularx}{\linewidth}{|l|l|X|}
\hline
\multicolumn{1}{|c|}{\textit{\textbf{Model}}} & \multicolumn{1}{|c|}{\textit{\textbf{Log Loss}}} & \multicolumn{1}{c|}{\textit{\textbf{Accuracy}}} \tabularnewline \hline
Random Forest & 2.312060 & 0.218282\tabularnewline \hline
Naive Bayes & 4.846123 & 0.274343\tabularnewline \hline
Decision Tree & 8.787213 & 0.322790\tabularnewline \hline
K Neighbors & 19.703055 & 0.195351\tabularnewline \hline
Logistic Regression & 9.2131214 & 0.052230\tabularnewline \hline
SVM & NaN & NaN\tabularnewline \hline
MLP & NaN & NaN\tabularnewline \hline
\end{tabularx}
\label{table:model_metrics}   
\end{table}

\section{Conclusion}
In this paper we have proposed a novel approach to predict multi-class crime type by incorporating unsupervised learning techniques and also relaxed some of the assumptions that we have seen in the current literature. We have kept working with all class labels and even though we got lower accuracy values, we were able to see that the best performing models were the same. When we  combine supervised and unsupervised learning techniques, our workflow also produced results that could be easily generalized to other cities, since we are not putting a grid on a city like other studies have done so far.  Due to lack of features and large number of class labels, we systematically crafted features in order to achieve better fit models. Specifically, we have described a methodology to run clustering algorithms on the data set, then use the distance to cluster centers as a feature in our supervised learning models. We achieved 2.2323 log loss on our Random Forest machine learning model, which was the best among various models that we have used. We hope this workflow of combining unsupervised and supervised learning models would give inspiration to create robust crime prediction workflows in fighting against crime.

\section{Acknowledgements}
We would like to acknowledge Dr. Andrew Cohen from Department of Electrical and Computer Engineering for teaching this course, Dr. Robert Kane from Department of Criminology and Justice Studies and Dr. Matthew Burlick from Department of Computer Science and Informatics for advising us.

\section{Future Work}
We have used Euclidean distance to calculate the distance from crime centers to crime points. Since the crimes are urban crimes, we would like to see the effect of choosing  a different distance such as city-block distance in the future work.

\end{document}


\title{Perfecting the Crime Machine \\
}

\section{Appendix}

Here we report more of our crime type aggregations over hours, months, and years.
We encourage the reader to explore these figures to see the underlying patterns for specific crime types.

\begin{figure}[!ht]
\centering
  \includegraphics[width=0.9\columnwidth]{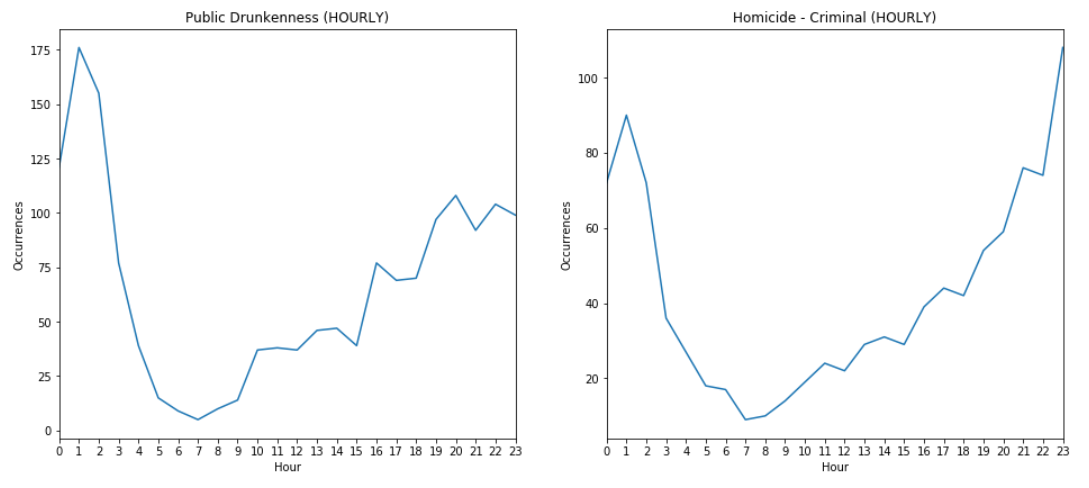}
  \caption{Public Drunkenness(i) and Homicide-Criminal(ii) crime counts aggregated over hours}~\label{fig:drunkness_hourly}
\end{figure}

\begin{figure}[!ht]
\centering
  \includegraphics[width=0.9\columnwidth]{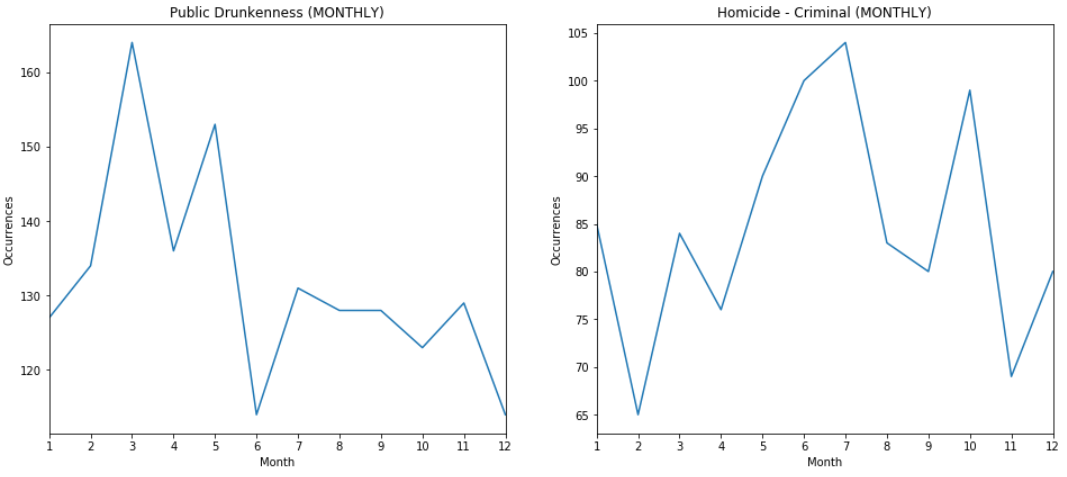}
  \caption{Public Drunkenness(i) and Homicide-Criminal(ii) crime counts aggregated over months}~\label{fig:drunkness_monthly}
\end{figure}

\begin{figure}[!ht]
\centering
  \includegraphics[width=0.9\columnwidth]{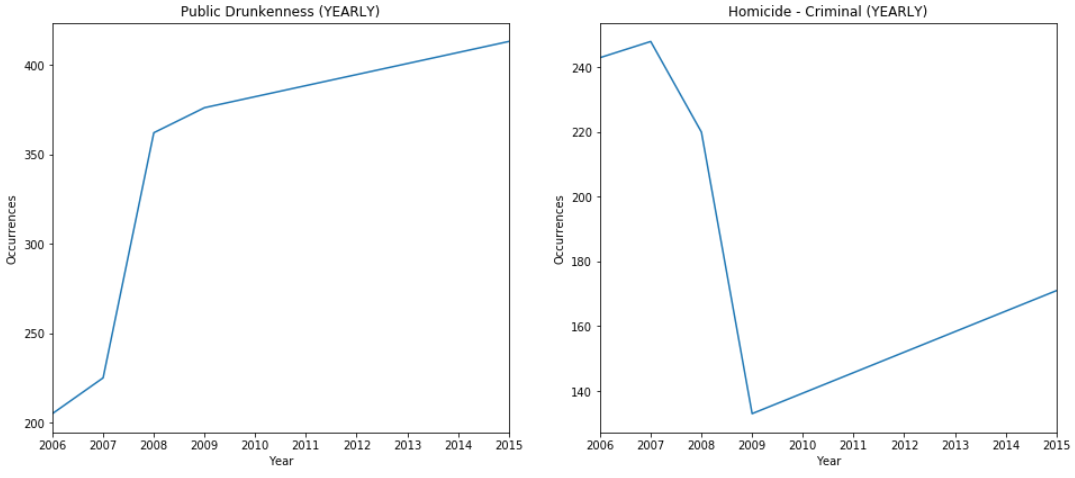}
  \caption{Public Drunkenness(i) and Homicide-Criminal(ii) crime counts aggregated over years}~\label{fig:drunkness_yearly}
\end{figure}

\begin{figure}[!ht]
\centering
  \includegraphics[width=0.9\columnwidth]{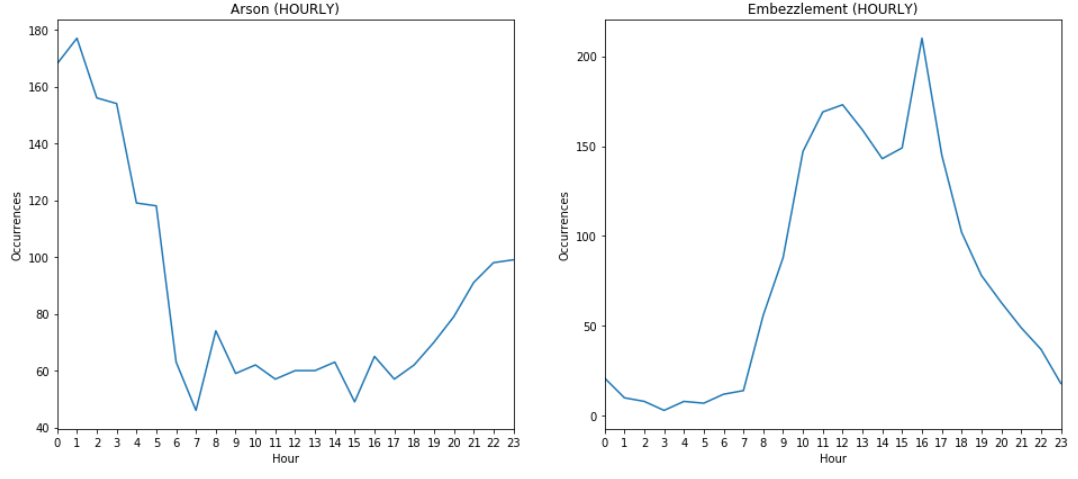}
  \caption{Arson(i) and Embezzlement(ii) crime counts aggregated over hours. We see that arson and embezzlement are total opposites from each other. }~\label{fig:arson_hourly}
\end{figure}

\begin{figure}[!ht]
\centering
  \includegraphics[width=0.9\columnwidth]{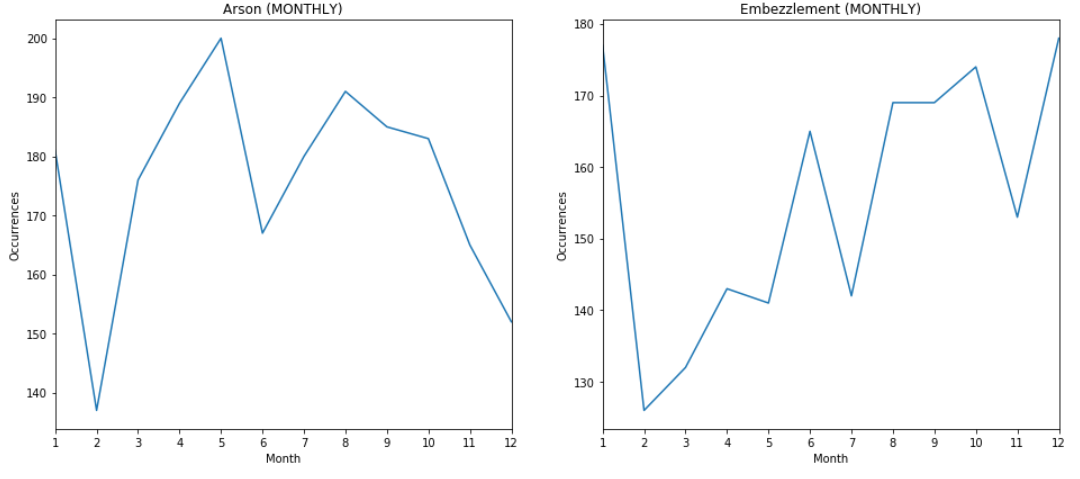}
  \caption{Arson(i) and Embezzlement(ii) crime counts aggregated over months. We see that cold months witness less crime incidents.}~\label{fig:arson_monthly}
\end{figure}

\begin{figure}[!ht]
\centering
  \includegraphics[width=0.9\columnwidth]{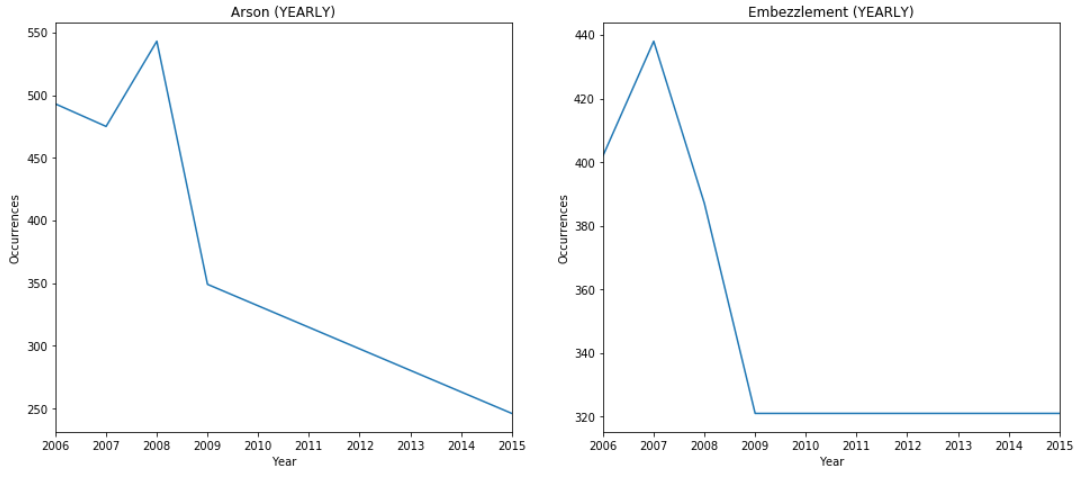}
  \caption{Arson(i) and Embezzlement(ii) crime counts aggregated over years. }~\label{fig:arson_yearly}
\end{figure}

\begin{figure}[!ht]
\centering
  \includegraphics[width=0.9\columnwidth]{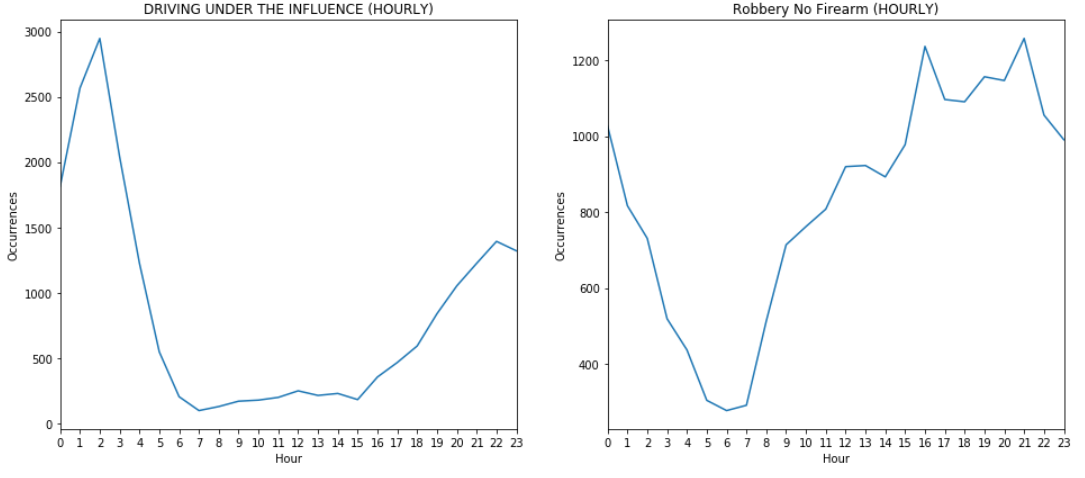}
  \caption{DUI(i) and Robbery No Firearm(ii) crime counts aggregated over hours}~\label{fig:dui_hourly}
\end{figure}

\begin{figure}[!ht]
\centering
  \includegraphics[width=0.9\columnwidth]{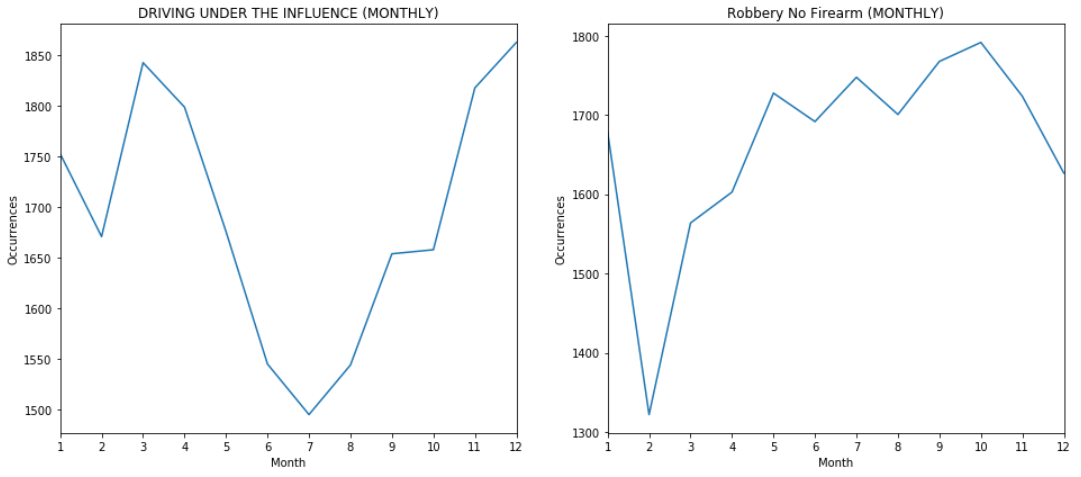}
  \caption{DUI(i) and Robbery No Firearm(ii) crime counts aggregated over months}~\label{fig:dui_monthly}
\end{figure}

\begin{figure}[!ht]
\centering
  \includegraphics[width=0.9\columnwidth]{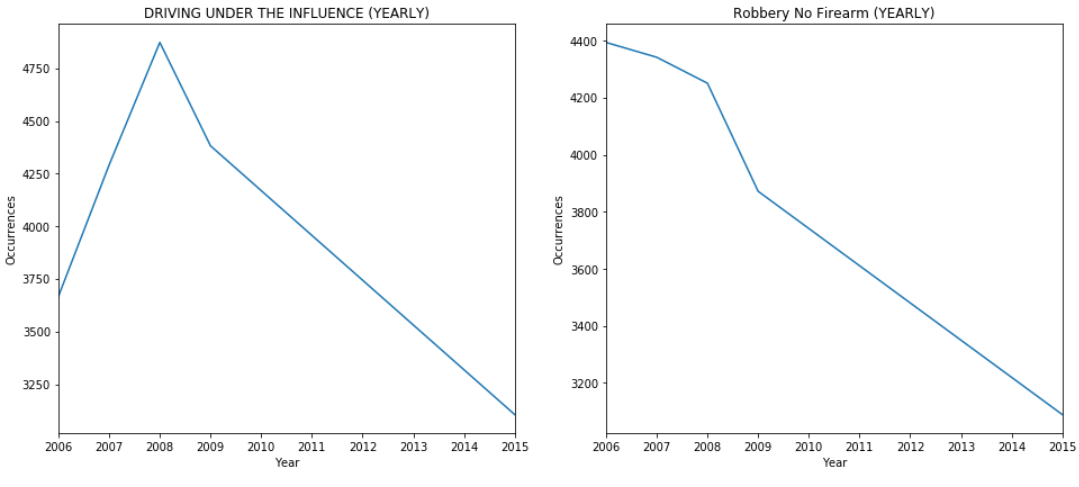}
  \caption{DUI(i) and Robbery No Firearm(ii) crime counts aggregated over years}~\label{fig:dui_yearly}
\end{figure}

\begin{figure}[!ht]
\centering
  \includegraphics[width=0.9\columnwidth]{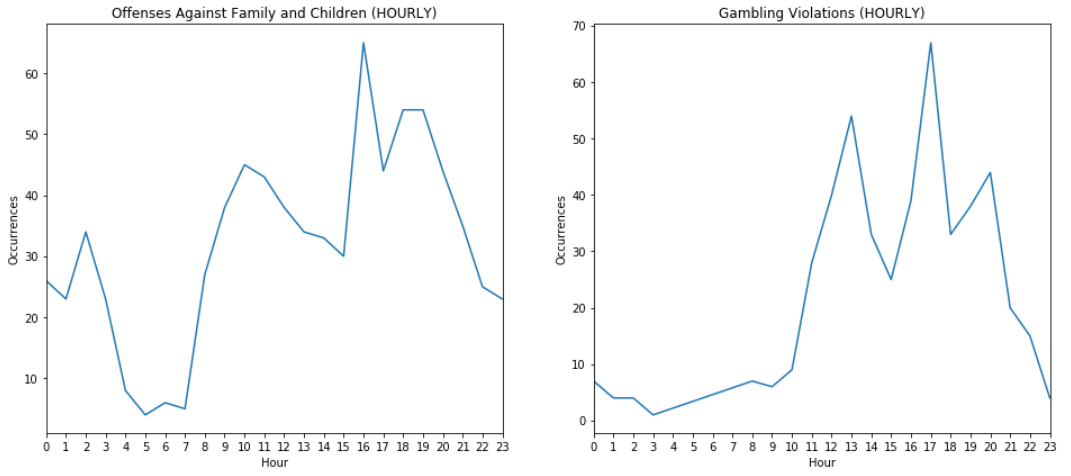}
  \caption{Offenses against family and children(i) and Gambling Violations(ii) crime counts aggregated over hours}~\label{fig:family_hourly}
\end{figure}

\begin{figure}[!ht]
\centering
  \includegraphics[width=0.9\columnwidth]{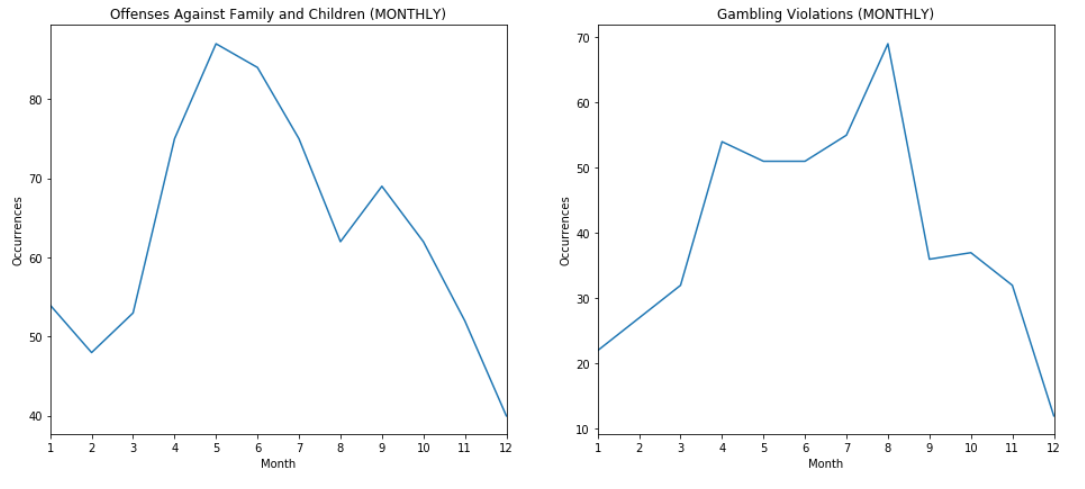}
  \caption{Offenses against family and children(i) and Gambling Violations(ii) crime counts aggregated over months}~\label{fig:family_monthly}
\end{figure}

\begin{figure}[!ht]
\centering
  \includegraphics[width=0.9\columnwidth]{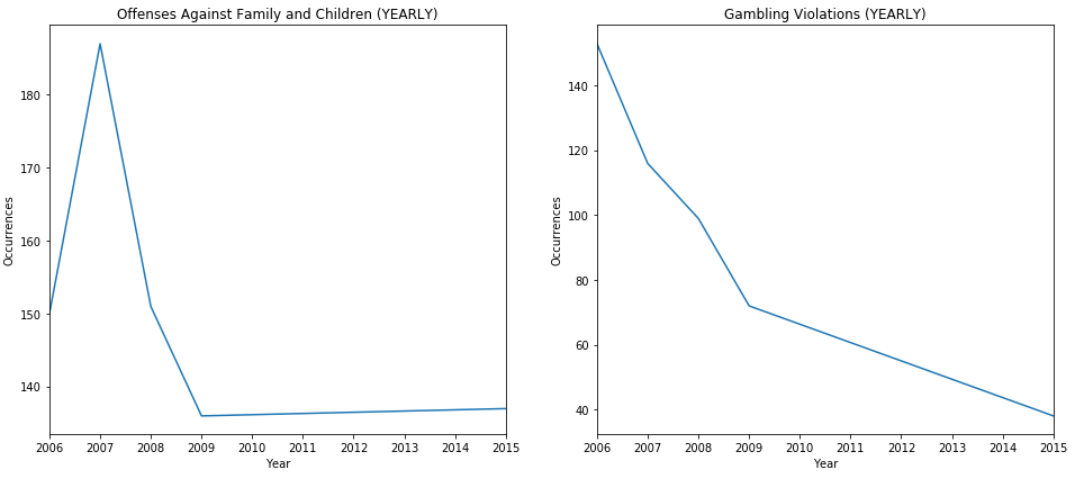}
  \caption{Offenses against family and children(i) and Gambling Violations(ii) crime counts aggregated over years}~\label{fig:family_yearly}
\end{figure}

\begin{figure}[!ht]
\centering
  \includegraphics[width=0.9\columnwidth]{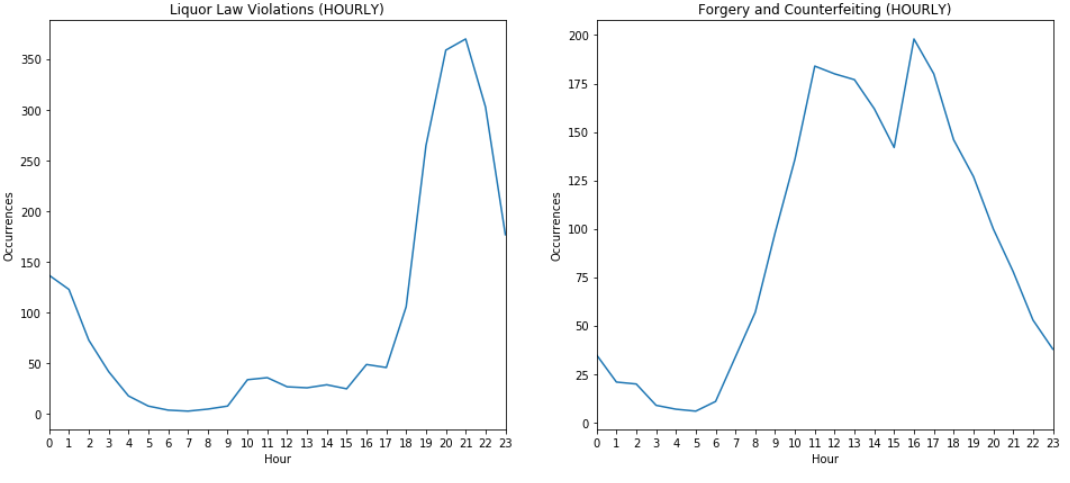}
  \caption{Liquor Law Violations(i) and Forgery and Counterfeiting(ii) crime counts aggregated over hours.}~\label{fig:forgery_hourly}
\end{figure}

\begin{figure}[!ht]
\centering
  \includegraphics[width=0.9\columnwidth]{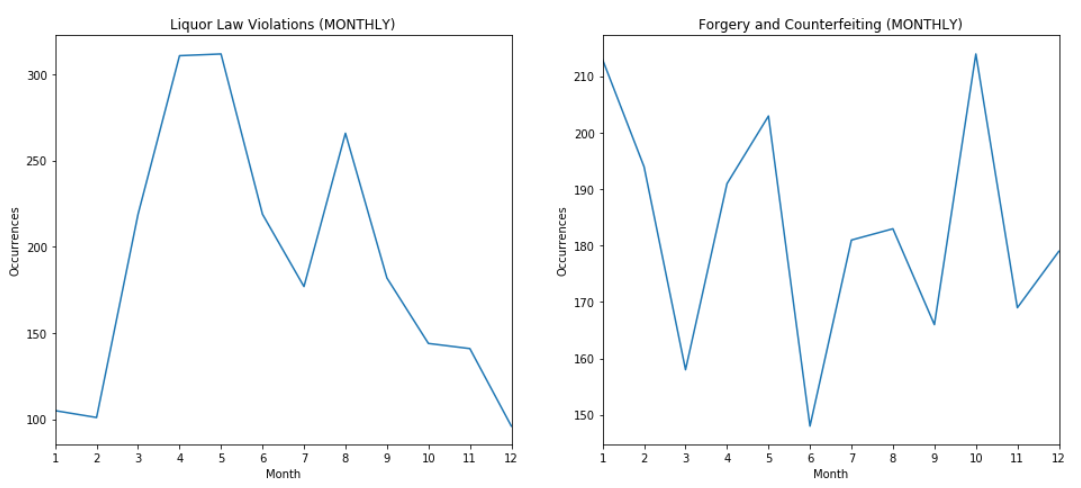}
  \caption{Liquor Law Violations(i) and Forgery and Counterfeiting(ii) crime counts aggregated over months.}~\label{fig:forgery_monthly}
\end{figure}

\begin{figure}[!ht]
\centering
  \includegraphics[width=0.9\columnwidth]{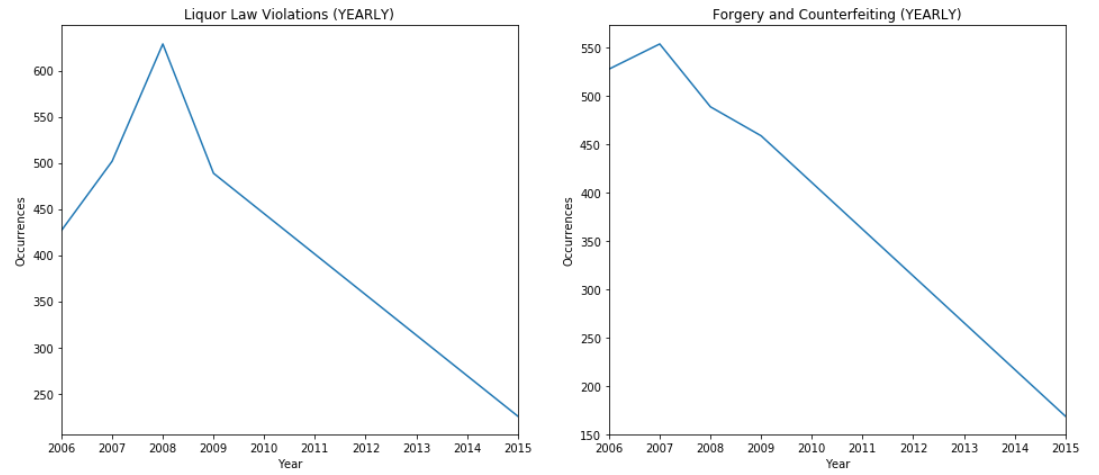}
  \caption{Liquor Law Violations(i) and Forgery and Counterfeiting(ii) crime counts aggregated over years.}~\label{fig:forgery_yearly}
\end{figure}

\begin{figure}[!ht]
\centering
  \includegraphics[width=0.9\columnwidth]{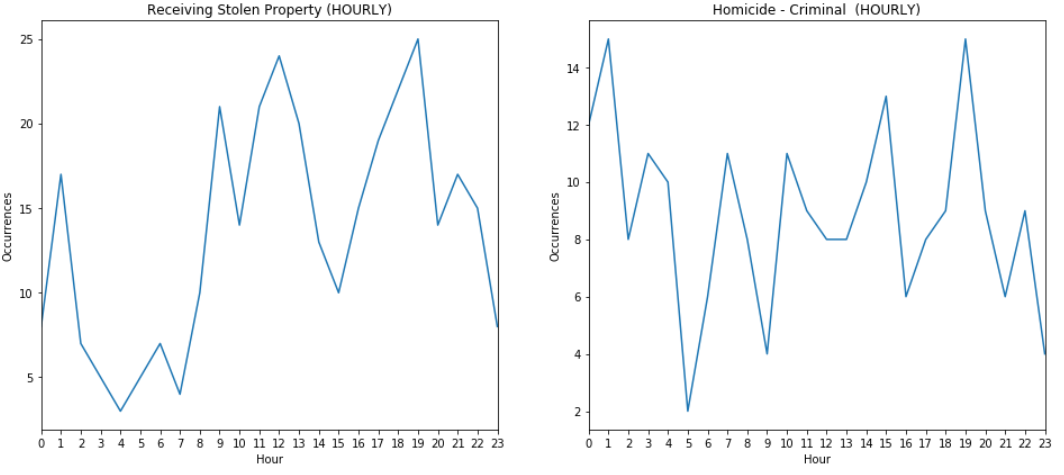}
  \caption{Receiving Stolen Property(i) Homicide-Criminal(ii) crime counts aggregated over hours.}~\label{fig:receiving_stolen_hourly}
\end{figure}

\begin{figure}[!ht]
\centering
  \includegraphics[width=0.9\columnwidth]{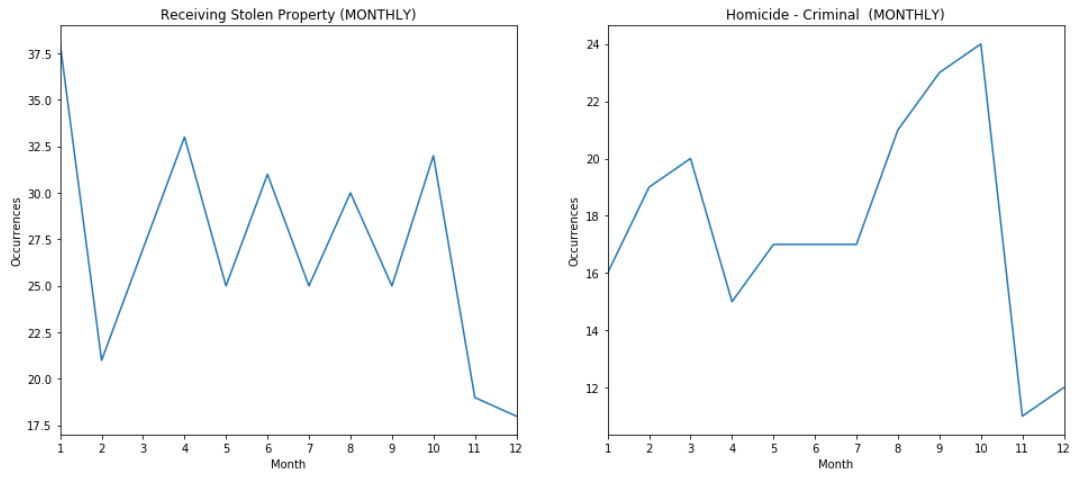}
  \caption{Receiving Stolen Property(i) Homicide-Criminal(ii) crime counts aggregated over months.}~\label{fig:receiving_stolen_monthly}
\end{figure}

\begin{figure}[!ht]
\centering
  \includegraphics[width=0.9\columnwidth]{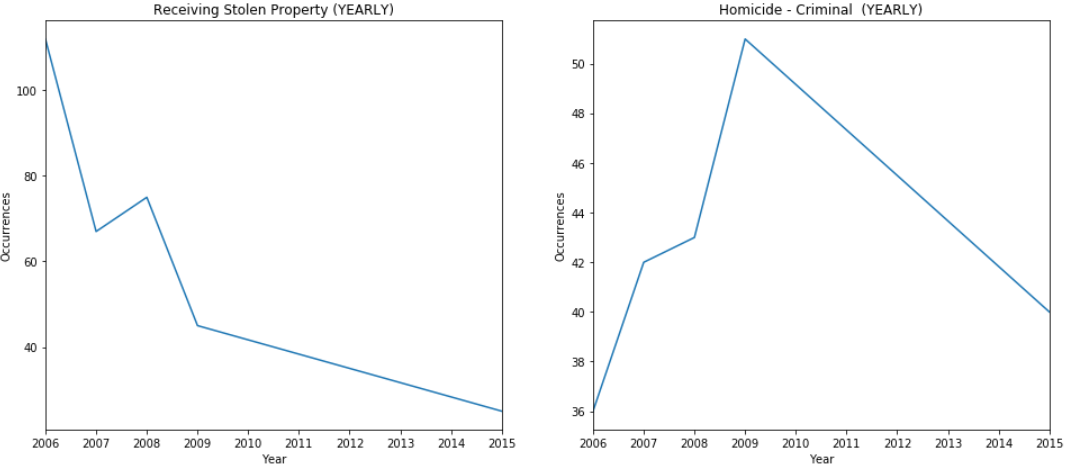}
  \caption{Receiving Stolen Property(i) Homicide-Criminal(ii) crime counts aggregated over years.}~\label{fig:receiving_stolen_yearly}
\end{figure}